\documentclass{vldb}
\usepackage{footmisc}
\usepackage{soul}
\usepackage{hyperref}
\usepackage{enumitem}
\usepackage{listings}
\usepackage{cleveref}
\usepackage{booktabs} %
\usepackage{xspace}
\usepackage{xcolor}
\usepackage{microtype}
\usepackage{subcaption}
\usepackage{balance}
\usepackage{mathtools}
\usepackage{scalerel,amssymb}
\usepackage{tcolorbox}%

\usepackage{adjustbox}

\newcommand{\key}[1]{\textbf{\textit{#1}}}

\newcommand{\myref}[1]{\S\ref{#1}\xspace}

\newcommand{\sys}{\textsc{Khameleon}\xspace}

\newcommand{\baseline}{\textsc{Baseline}\xspace}
\newcommand{\pred}{\textsc{Predictor}\xspace}
\newcommand{\prog}{\textsc{Progressive}\xspace}
\newcommand{\kalman}{\textsc{Kalman}\xspace}
\newcommand{\onhover}{\textsc{OnHover}\xspace}
\newcommand{\uniform}{\textsc{Uniform}\xspace}
\newcommand{\oracle}{\textsc{Oracle}\xspace}

\newcommand{\smalldb}{\textsc{Small}\xspace}
\newcommand{\bigdb}{\textsc{Big}\xspace}

\definecolor{darkgreen}{rgb}{0.15,0.55,0.15}
\definecolor{darkblue}{rgb}{0.1,0.1,0.5}
\definecolor{blue}{rgb}{0.01,0.40,.8}
\definecolor{darkgreen}{rgb}{0.15,0.55,0.15}
\definecolor{mred}{rgb}{.80,.12,.30}
\definecolor{grey}{rgb}{0.5,0.5,0.5}
\definecolor{Purple}{rgb}{.75,0,.85}
\definecolor{light-gray}{gray}{0.95}
\definecolor{mid-gray}{gray}{0.85}
\definecolor{darkred}{rgb}{0.7,0.25,0.25}
\definecolor{rose}{rgb}{1.0, 0.01, 0.24}
\newcommand{\red}[1]{\textcolor{red}{#1}}
\newcommand{\grey}[1]{\textcolor{grey}{#1}}

\newcommand{\blue}[1]{\textcolor{blue}{#1}}

\newtcbox{\redbox}{on line,
  colframe=white,colback=red!10!white,
  height=1em,valign=bottom,
  boxrule=0.5pt,arc=2pt,boxsep=0pt,left=2pt,right=2pt,top=1pt,bottom=1pt}
\newtcbox{\bluebox}{on line,
  colframe=white,colback=blue!10!white,
  height=1em,valign=bottom,
  boxrule=0.5pt,arc=2pt,boxsep=0pt,left=2pt,right=2pt,top=1pt,bottom=1pt}

\newcommand{\eat}[1]{}

\newcommand{\stitle}[1]{\smallskip\noindent\textbf{#1}}

\newlength{\listingindent}                %
\usepackage[LGRgreek]{mathastext}

\DeclarePairedDelimiter\floor{\lfloor}{\rfloor}

\newtheorem{thm}{Theorem}

\newtheorem{problem}[thm]{Problem}

\begin{document}
\setcounter{page}{1}
\setcounter{section}{0}
\pagenumbering{arabic}

\title{Continuous Prefetch for Interactive Data Applications}
\subtitle{Technical Report
\thanks{this is an extended version of a paper that appeared in \textit{VLDB 2020}.}
}

\def\refColumbia{\raisebox{4pt}{$\ddagger$}}
\def\refUCLA{\raisebox{4pt}{$\dagger$}}
\def\refCornell{\raisebox{4pt}{$\S$}}

\numberofauthors{1}
\author{
    Haneen Mohammed\refColumbia{}, 
    Ziyun Wei\refCornell{}, 
    Eugene Wu\refColumbia{}, 
    Ravi Netravali\refUCLA{}
    \\
    \refColumbia{}Columbia University,
    \refCornell{}Cornell University,
    \refUCLA{}UCLA
}

\maketitle

\begin{abstract}
Interactive data visualization and exploration (DVE) applications are often
network-bottlenecked due to bursty request patterns, large response
sizes, and heterogeneous deployments over a range of networks and devices.
This makes it difficult to ensure consistently low response times
($<100ms$).  \sys is a framework for DVE applications that
uses a novel combination of prefetching and response tuning to dynamically
trade-off response quality for low latency.

\sys exploits DVE's approximation
tolerance: immediate lower-quality responses are preferable to waiting for
complete results.  To this end, \sys progressively encodes responses, and runs
a server-side scheduler that proactively streams portions of responses using
available bandwidth to maximize user-perceived interactivity.  The scheduler involves
a complex optimization based on available resources, predicted user
interactions, and response quality levels; yet, decisions must also be
made in real-time. To overcome this, \sys uses a fast greedy heuristic that
closely approximates the optimal approach.  Using image exploration and
visualization applications with real user interaction traces, we show that
across a wide range of network and client resource conditions, Khameleon
outperforms existing prefetching approaches that benefit from \emph{perfect}
prediction models: \sys always lowers response latencies (typically by 2--3
orders of magnitude) while keeping response quality within $50$--$80\%$.

\end{abstract}

\section{Introduction}
\label{s:intro}

\begin{figure}
  \centering
  \begin{subfigure}[t]{.95\columnwidth}
	\centering
	\includegraphics[width = .95\columnwidth]{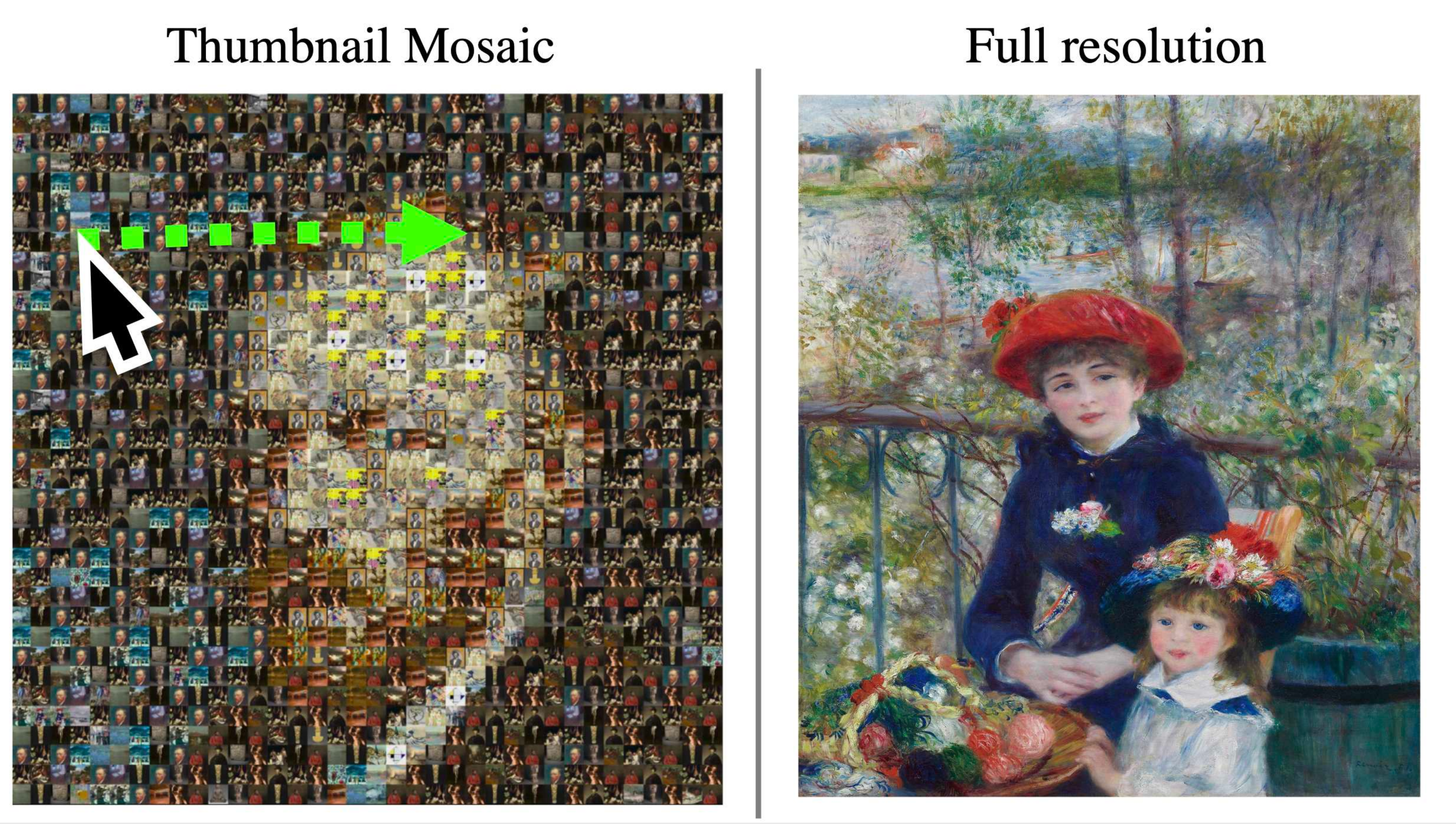}
	\caption{ }
	\label{f:vis_mosaic}
  \end{subfigure}\hfill\\
   \begin{subfigure}[t]{.95\columnwidth}
 	\centering
 	\includegraphics[width =  .95\textwidth]{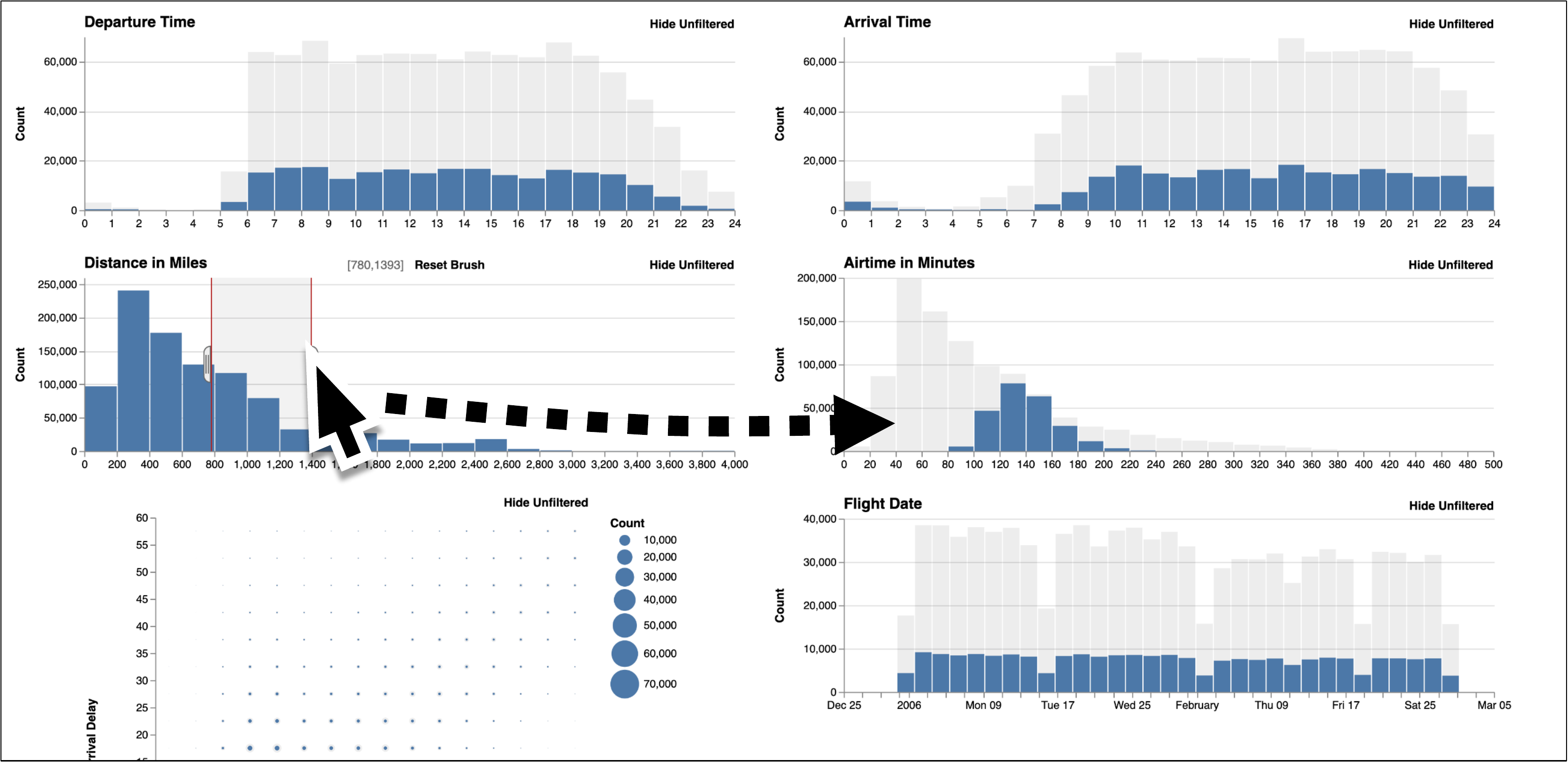}
    \caption{ }
 	\label{f:vis_falcon}
   \end{subfigure}
  \vspace{.1em}
  \caption{Example interactive DVE applications. (a) Image exploration application; hovering over the mosaic of thumbnails on the left loads the high resolution image on the right. (b) Falcon interactive visualization~\protect\cite{moritz2019falcon}; users drag and resize range filters on any subset of charts, which triggers updates to the others. We evaluate these applications with and without \sys in \myref{s:experiments}. }
  \label{f:vis}
\end{figure}

Interactive data visualization and exploration (DVE) applications, such as
those in \Cref{f:vis}, are increasingly popular and used across sectors
including art galleries~\cite{chicagoart}, earth science~\cite{landsat},
medicine~\cite{cellatlas}, finance~\cite{mapdimmerse}, and
security~\cite{graphistry}.   Like typical web services, DVE applications may
be run on heterogeneous client devices and networks, with users expecting fast
response times under 100
ms~\cite{Deber2015HowMF,liu2014effects,Rahman2019EvaluatingID}. However, the
resource demands of DVE applications are considerably magnified and highly
unpredictable, making it difficult to achieve such interactivity.

Traditional interactive applications are based on point-and-click interfaces
such as forms or buttons, where there may be seconds or minutes of delay
between user requests.  In contrast, DVE applications update the visual
interface continuously as the user drags, hovers, or otherwise manipulates the
interface~\cite{Dimara2020WhatII} (\Cref{f:vis}). For example, all of the charts in \Cref{f:vis_falcon}
are updated continuously as the user drags and resizes range filters.
In short, DVE applications support a large number of potential requests rather than a few buttons,
exhibit \key{bursty}~\cite{Battle2019CharacterizingEV} user interactions that
generate a huge number of back-to-back requests with nearly \key{no ``think time''} between them,
and issue \key{data-dense} requests for tens of kilobytes to megabytes of data 
in order to render detailed statistics or high-resolution images~\cite{CruzNeira1993SurroundscreenPV}.

As a result of these combined factors, DVE applications place considerable
and unpredictable pressure on both network and server-side data processing
resources.  Fortunately, the database and visualization communities have made
considerable progress in reducing
server-side~\cite{mapdimmerse,Zukowski2012VectorwiseBC,liu2013immens,chen2006cgmolap,wu2019anna}
and client-side~\cite{graphistry,liu2013immens} data processing and rendering
latencies. However, {\it network bottlenecks still persist}, and can
cripple user-facing responsiveness even if server and client-side overheads are
eliminated.  Addressing network bottlenecks is becoming paramount with the continued shift towards cloud-based DVE applications that must remain
responsive across a wide variety of client network conditions (e.g., wireless,
throttled).

The primary approach to masking network delays in interactive applications is
prefetching~\cite{doshi2003prefetching,battle2016dynamic,dice2014,domenech2006web,white2017search,chan2008maintaining,ahmed2012adaptive},
where responses for predicted future requests are proactively cached on the
client before the user requests them. Prefetching benefits inherently depend on
prediction accuracies, but sufficiently high accuracies have remained elusive, even for
well-studied classic applications such as web pages~\cite{lenin_prefetch}.
DVE applications pose an even more challenging setting for several reasons. 
Their bursty request patterns, when combined with data-dense
responses, can easily exceed the bandwidth capacities of existing networks and
cause persistent congestion.  At the same time, the massive number of potential requests
makes building a near-perfect predictor that can operate over long time
horizons infeasible---developing such oracles is an open problem.  Thus,
prefetching for DVE applications is either ineffective or wastes precious
network resources which, in turn, can cause (detrimental) {\it cascading slowdowns} on later
user requests.

In this paper, we depart from traditional prefetching frameworks that hope to
accurately predict a small number of future requests to prefetch, towards a framework that continuously and aggressively hedges across a large set of
potential future requests.  Such framework should also decouple request
burstiness from resource utilization so that the network does not get
overwhelmed at unpredictable intervals, and instead can consistently employ all
available network resources for subsequent prefetching decisions.

A trivial, but undesirable, way to meet these goals is to limit the user's ability to interact with the interface, thereby reducing the burstiness and scope of possible requests.
Instead, we leverage the fact that DVE applications are
\key{approximation tolerant}, meaning that response quality can be dynamically tuned~\cite{Mao2017NeuralAV,Yin2015ACA,geotiff} to enable more hedging within
the available resources (at the expense of lower response quality).  Of course,
this introduces a fundamental tradeoff: the prefetching system can focus on
low-quality responses for many requests to ensure immediate responses, or
high-quality responses for a few requests at the risk of more cache misses and
slow responses.  Balancing this tradeoff requires a \textit{joint optimization between
response tuning and prefetching}, which, to date, have only been studied
independently.  This involves a novel and challenging scheduling problem, as the optimization space 
needs to consider the likelihood of the user's future access patterns over a large number of possible requests,
applications preferences between response quality levels and responsiveness, and limited resource conditions.
At the same time, the scheduler must run in real-time.

We present \textbf{\sys}, a novel prefetching framework for DVE applications
that are bottlenecked by request latency and network transfer.  \sys
dynamically trades off response quality for low latency by leveraging two
mechanisms that collectively overcome the aforementioned joint optimization
challenges.

First, we leverage progressive encoding\footnote{Progressive encoding is distinct from progressive computation, 
such as online aggregation~\cite{hellerstein1997online}, which returns full, yet approximate, responses by processing a sample of the database.  
\myref{s:related} discusses how progressive computation can exacerbate network bottlenecks in more detail.\label{fnlabel}} to enable fine-grained scheduling.  
Each response is encoded as an ordered list of small blocks such that any prefix is sufficient to
render a lower quality response, and additional blocks improve response quality.  
This encoding is natural for DVE applications,
which can leverage existing encodings for e.g., 
images~\cite{wallace1992jpeg,geotiff} and visualization data~\cite{battle2016dynamic}. 
Progressive encoding lets the prefetching system 
vary the prefix for a given response based on its (predicted) future utility to the user.

Second, we shield network and server resources from prediction errors and client burstiness by using
a push-based model, rather than having clients issue requests directly to the server.
The server streams blocks for likely requests to the client cache using the
available (or user-configured) network capacity.  The server-side scheduler
determines a global sequence of blocks to continually push to the client.  By
default, it assumes that all requests are equally likely.  However, the application can
define a predictor to estimate future requests; in this case, the client uses the predictor
to periodically send forecasted probability distributions to the scheduler,
which updates the global sequence accordingly. To ensure real-time operation,
the scheduler uses a greedy heuristic, which closely approximates the optimal
algorithm.

\sys is a framework that is compatible with existing DVE applications. \sys
transparently manages the request-oriented communication between the DVE client
and server, and shields developers from the challenges of the joint
optimization problem.  Developers can instead focus on high-level policies,
such as determining their preference between latency and quality, and
developing application-specific progressive encoding schemes and prediction
models.  \myref{s:arch-adapt} describes the steps that a developer
would take to use \sys with an existing DVE application.

We evaluate \sys using the two representative DVE applications in \Cref{f:vis}.
Our experiments consider a broad range of network and client resource
conditions, and use real user-generated interaction traces.  Across these
conditions, we find that \sys is able to avoid network congestion and
degraded user-facing responsiveness that arises from using indiscriminate
prefetching (even if that prefetching uses a 100\% accurate predictor).  For instance, for
the image exploration application, \sys (using a simple
predictor~\cite{welch1995introduction}) reduces response latencies by up to 3
orders of magnitude ($>10s$ to $\approx 10ms$) and maintains a response quality
of $50$--$80\%$. Similarly, with the Falcon data
visualization~\cite{moritz2019falcon}, \sys's progressive encoding improves
response latencies on average by
$4\times$
and improves response quality by up to
$1.6\times$. Our experimental setup also reveals that porting existing applications to use
\sys entails minimal burden. For example, modifying Falcon to use \sys as
the communication and prefetching layer required fewer than 100 lines of code
to issue requests to the \sys client library and use a formal predictor.

To summarize, our contributions include 1) the design and implementation of
\sys, a framework that combines real-time prediction, progressive encoding, and
server-side scheduling to address the diverse challenges of interactive DVE
applications,  2) the formalization of the server-side scheduling optimization
problem that explicitly balances the quality and likelihood of a request, along
with a fast greedy heuristic implementation, 3) and an extensive evaluation
using two interactive applications that highlight the benefits of the \sys
design.

\pagebreak
\section{DVE Applications}
\label{s:usecases}

Cloud-based DVE applications are \key{information dense}, in that they render hundreds or
thousands of data items (e.g., records, images) that users can directly interact with.  The request
patterns from these interactions are \key{bursty} with \key{negligible think time} between
requests.  These characteristics lead to a rate of requests that stresses the
network, often exceeding the available capacity and resulting in congestion.

In order to address the potential network bottlenecks, \sys leverages two key
properties of DVE applications. First, interactions are \key{preemptive}: since
responses can arrive out of order (e.g., due to network or server delays), the
client renders the data for the most recent request and (silently) drops
responses from older requests to avoid confusing the
user~\cite{Wu2016ADA,Wu2018MakingSO}.  Second, they are \key{approximation
tolerant}: it is preferable to quickly render a low-quality response (e.g.,
fewer points~\cite{rahman2017ve} or coarser bins~\cite{liu2013immens}) than to
wait for a full quality response. As concrete examples, consider the following
two DVE applications which exhibit these properties; we use both in our
evaluation (\myref{s:experiments}).

\stitle{Large-scale image exploration.} 
Scientists and users increasingly wish to interactively explore massive image
datasets of e.g., art~\cite{chicagoart}, satellite data~\cite{landsat}, cellular
microscopy~\cite{cellatlas}, and maps~\cite{Haklay2008OpenStreetMapUS}.  Along
these lines, we developed an image gallery DVE application (\Cref{f:vis_mosaic}).  The user's mouse hovers over
a dense array of $10,000$ image thumbnails on the left (akin to a
zoomed-out view) to view the full resolution $1.3-2Mb$ image of the
hovered-over thumbnail on the right (akin to a zoomed-in tile).

We consider this an exemplar and difficult DVE application to evaluate, because it has a high
request rate, large response sizes, and with 10K thumbnails, it is difficult to
build an accurate predictor for. For instance, from the user traces used in our
experiments, clients request up to 32 images per second (32--64
megabytes(MB)/s),\footnote{For reference, streaming HD and 4K video typically
requires 5-20 Megabits (Mb)/s.} not including any prefetching requests.
In addition, this application imposes fewer interaction restrictions than existing exploration
applications that are plagued by prefetching inefficiencies.
For instance, applications like Google Maps only let users pan to adjacent tiles
and incrementally zoom; this both simplifies prediction and limits the rate of
images that the user can request at a time.

\stitle{Interactive data visualizations.}
Falcon~\cite{moritz2019falcon} is a state-of-the-art interactive visualization application
specifically optimized for prefetching (\Cref{f:vis_falcon}).
As the user selects and resizes range selections in any of the charts, the other non-selected charts
immediately update their statistics to reflect the conjunction of all of the
selections.  The challenge is that the space of possible combinations of
selections is exponential in the number of charts, and is infeasible to fully
precompute and send to the client up front. Yet, even movements across a single pixel
trigger many requests to update the charts.  

In order to minimize interaction delays, the Falcon developers~\cite{moritz2019falcon} manually implemented prefetching
to mask request latencies.  They observed that the user can only interact with
one chart at a time, and in the meantime, selections in the other charts are fixed. 
When the user's mouse moves onto chart A, Falcon sends SQL queries to a backend
database to compute low dimensional data cube slices between chart A and each of the other
charts to update.  Once these slices are constructed, user interactions in
chart A are handled instantaneously.

Falcon's {\it predictor} prefetches data slices when
the user hovers over a chart, and it {\it progressively encodes} the data
slices as cumulative counts.  However, these
policies are hardcoded in a monolithic codebase, making it challenging
to improve the predictor (e.g., estimate the chart the mouse will
interact with, rather than wait for a hover event), response encoding
(e.g., pixel-resolution and a coarse resolution ), or user preferences (e.g., which attributes they favor).
\myref{ss:exp_falcon} describes the details of how we adapted Falcon to use \sys as the communication layer, and switched its database from OmniSci to PostgreSQL.

\section{Khameleon Overview}\label{s:arch}

\begin{figure*}
  \centering
  \includegraphics[width=\textwidth]{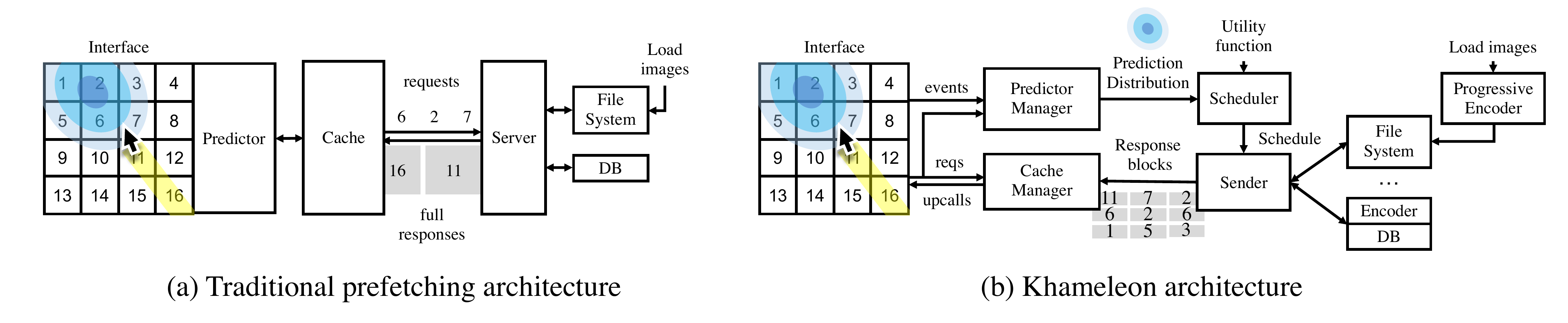}
  \caption{Comparing \sys to a traditional prefetching architecture for an image exploration DVE application. The interface is a $4\times4$ grid of image thumbnails. The cursor moved over images 16 and 11 (yellow line) and is now positioned over image 7; the probability of the mouse's future position is a gaussian distribution illustrated by the blue ellipses (center is the highest probability).  The gray boxes are sized according to response sizes.  \sys separates the predictor from the cache, sends probability distributions instead of explicit requests, and uses a scheduler to determine the sequence of small request blocks to send to the client. }
  \label{fig:arch}
\end{figure*}

This section first describes traditional prefetching architectures and their
limitations for DVE applications. It then provides a high-level design overview
of \sys, explaining its individual components (we elaborate on each in the
following sections), how they collectively overcome the limitations of prior
architectures, and how existing DVE applications can be seamlessly migrated to
use \sys.

\subsection{Traditional Prefetching Architecture}
\label{s:arch-traditional}

  \Cref{fig:arch}(a) depicts the workflow of a common prefetching architecture
  for an image exploration DVE application. In this application, a user interacts
  with a grid composed of 16 image thumbnails such that mousing over an image
  enlarges its display size. 
  As the user moves the mouse, the local cache manager receives requests and
  immediately responds if the data is cached, or forwards the request to the
  server.  In parallel, the gaussian distribution representing predictions of the
  mouse's future location is updated based on the mouse's current position (to
  improve prediction accuracy), and is used to pick a set of requests
  to prefetch; these requests are issued to the cache manager in the same way.

  In this example, the user mouse has moved quickly along the yellow path,
  triggering requests for images 16 and 11. Given the bursty nature of these
  requests, the corresponding responses are still being transmitted as the next
  user-generated request is issued (for image 7). Unfortunately, because the full
  response data for images 16 and 11 fully utilize the available network
  bandwidth, the response for 7 will have to contend with those old responses,
  delaying its arrival. To make matters worse, the client prefetching mechanism
  will concurrently issue requests for the $k$ most likely next requests (2 and 6
  in this example).

  \stitle{Limitations.} The problem here is that the information that enables the most accurate
  predictions (i.e., the mouse's current position) is available exactly when the
  user issues bursts of requests that already congest the network. This has two
  consequences. First, this largely eliminates any prefetching benefits, and
  highlights its drawbacks: accurate prefetching requests are unlikely to return
  to the cache before the user explicitly requests them (recall that DVE
  applications experience low user think times), and inaccurate prefetching
  requests add unnecessary network congestion that slows down explicit
  user-generated requests and future prefetching. Second, it is difficult to know what
  requests should be prefetched during the times between user interactions because 
  the user, by definition, is not generating events;
  unfortunately, prefetching everything is impractical given the high
  data footprint of DVE applications (\S\ref{s:usecases}).

\subsection{Khameleon Architecture}

\sys (\Cref{fig:arch}(b)) consists of client-side and server-side libraries
that a cloud-based DVE application can import and use to manage data-intensive
network communication. These components operate as follows to overcome the
aforementioned limitations of traditional prefetching architectures.

The client-side library serves to decouple prefetching requests from the (bursty) network utilization
triggered explicitly by the user. User-generated
requests are not sent out on the network, and instead are registered with the
local {\it Cache Manager}. The Cache Manager waits until there is
cached data to fulfill the request, and then makes an application upcall to
update the interface with that data. This approach helps absorb high user
request rates. As this happens, client events (e.g., mouse movements) and
requests are also passed to an application-provided {\it Predictor Manager} that
continually updates a distribution of predicted future requests and sends a
summary of that distribution (e.g., the parameters of a gaussian distribution)
to the server.

The server-side library uses intelligent push-based scheduling and progressive
encoding of responses to make the most of the available network resources,
i.e., balancing user-perceived latency and response utility while hedging
across potential future requests. The \emph{Scheduler} continually maintains a
schedule of response blocks to push to the client; the set of blocks covers a
wide range of explicit and anticipated requests, e.g., images 11, 7, etc. in
Figure~\ref{fig:arch}(b).  The specific sequence of blocks depends on the
predicted request probabilities received from the client, as well as an
optional application-provided \emph{Utility Function} that quantifies the
``quality'' of a response based on the number of prefix blocks available. Note
that a single block is a complete response, with additional blocks improving
``quality'' according to the Utility Function.  A separate \emph{Sender} thread
reads the schedule and retrieves blocks from backend systems.  For example, the
file system could be pre-loaded with the blocks for progressively encoded
images, or a database could dynamically execute queries and progressively
encode the results before returning the subset of required blocks to the
Sender.  Finally, the server streams the sequence of response blocks to the
client, which updates its local cache accordingly.

As we describe in \myref{s:arch-adapt}, this design enables the application to
independently improve its prediction model, utility functions, data encodings,
backends, or scheduling policies. The \sys architecture is agnostic to how the
application client interprets and decodes the blocks, as well as to the specific
backend system that is employed.

\subsection{System Components}\label{ss:archcomponents}

\stitle{Predictor Manager.}  This client-side component relies on an
application-provided predictor to forecast future requests (as a probability
distribution over the possible requests), and periodically sends those
predictions to the server.  Predictors must satisfy two properties.  First, at
time $t$, the predictor must return a probability distribution $P^t(q|\Delta)$
over requests $q$ and future times $t+\Delta$.   Second, it must be {\it
Anytime}, so that the Predictor Manager can ask for distributions of predicted
requests to send the server at any time during system operation.  It is also
important that the predictor's state is efficient to maintain, and that the
distributions can be compactly represented for transmission to the server.  These
mechanisms enable the Predictor Manager to control policies for how often to
make and send distributions.

\stitle{Progressive Results and Caching.} Each request's progressively encoded
response is modeled as an ordered list of fixed size blocks;  any prefix
is sufficient to render (a possibly lower quality) result, and the full set of
blocks renders the complete result.  Smaller blocks can be padded if block sizes differ.  
Our client-side cache implementation uses
a ring buffer (FIFO replacement policy) for its simplicity and determinism; in
particular, this simplifies the server-side scheduler's ability to track cached
state at the client, since the FIFO policy can be simulated without explicit
coordination.\footnote{Other deterministic replacement policies are
possible and incorporating them is left for future work.}

During operation, the cache puts the $i^{th}$ block received from the server into slot $i\%C$,
where $C$ is the cache size.   The cache responds to a request if there is
$\ge1$ response block in the cache for the corresponding request.  To implement
preemptive interactions (\myref{s:usecases}), the cache assigns each request an
increasing logical timestamp when it is registered, and deregisters all earlier
timestamps when an upcall for request $i$ is made.

\begin{figure}[t]
  \centering
  \includegraphics[width = 0.7\columnwidth]{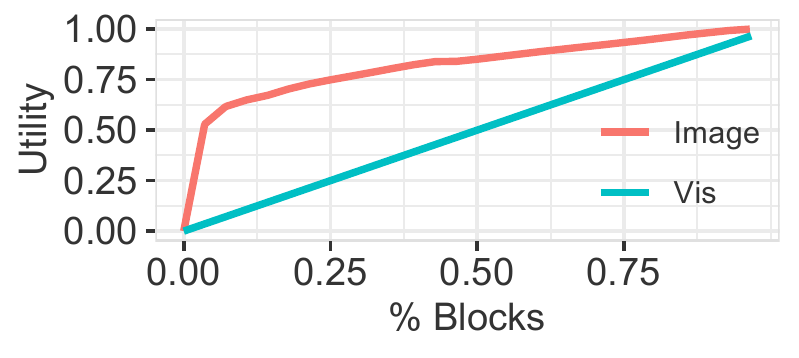}
  \caption{\small Utility function for image exploration application using structural similarity (red) and the visualization application which uses the system default linear function (blue).}
  \label{f:utility}
\end{figure}

\stitle{Utility Functions.} In practice, the first few blocks of a
response are likely to contribute more than subsequent blocks~\cite{rahman2017ve,wallace1992jpeg,malvar1999fast}.
To quantify this, the application can optionally provide a monotonically increasing {\it
Utility Function} $U:[0, 1]\mapsto [0, 1]$, which maps the percentage of data
blocks for a request to a utility score.  A score of 0 means most dissimilar,
and 1 means identical to the full result, in expectation. By
default, \sys conservatively assumes a linear utility function. As
an example, \Cref{f:utility} plots the utility curve for the image exploration
application, which is based on the average visual structural similarity
measure~\cite{Wang2004ImageQA} between the progressive result and the full
image.

\stitle{Scheduler and Backends.} The Scheduler decides the sequence of blocks
to push to the client.  It schedules in batches of $C$ blocks (the client
cache size) because the client's ring buffer will overwrite itself once it is
filled.  A given batch (a {\it schedule}) is chosen to maximize the user's expected utility with respect
to the probability distribution over future requests.  The separate {\it
Sender} thread reads the schedule to learn about the order in which blocks
should be retrieved from the backend and placed onto the network.  The backend
may be a file system, a database engine, a connection pool, or any service that
can process requests and return progressively encoded blocks. Note that, given a progressive encoder, any backend can be retrofitted by encoding its results. We retrofit PostgreSQL for our visualization experiments.

By default, we assume that retrieving blocks from the backend incurs a
predictable delay.  In addition, we assume that the backend is {\it
scalable}, in that the delay does not increase considerably when more
concurrent queries are issued (e.g., speculatively for prefetching).  This is
natural for precomputed responses or backends such as a file system or key
value store.  In cases where the backend can only scale to a limited number of
requests, \sys employs a heuristic to limit the amount of speculation in accordance with
the supported scalability (\myref{ss:impl}).

\subsection{Adapting Applications to Khameleon}
\label{s:arch-adapt}

  This subsection describes how a DVE application (image exploration in this case) can be easily and incrementally adapted to use \sys.  
Recall that the application issues an image request when the user's mouse hovers over a thumbnail; the server retrieves the full-sized image from the file system, and sends it back to the client.  

  To use \sys, the application should provide a progressive encoding of its responses, a utility function, and a predictor.  
Since traditional requests and responses are special cases of \sys's predictor and encoder, we start with generic defaults for these components.
 The generic encoder treats each image as a response with a single block, and the  predictor treats each request as a point distribution.  By specifying this, an immediate benefit is that the scheduler will use the point distributions to select the full requested image (as in the existing application), and use the remaining bandwidth to push random images for the client to cache.   
 
We now show how a developer Jude can improve each component for her application.  A benefit of \sys's modular  design is that the
components can be improved independently.

\stitle{Improve the Encoder:} Finer-grained blocks help improve the scheduler's ability to hedge across many requests given finite bandwidth resources.  Since JPEG is a progressive encoding, Jude replaces the naive encoder with a JPEG encoder and configures the system with the block size.   Further, she can adjust the JPEG encoding parameters to create finer-grained block sizes, or switch to an alternative progressive encoding altogether~\cite{pintus2011objective}.

  \stitle{Improve the Utility Function:} By default, \sys uses the linear utility function, where each block contributes the same additional utility to the user.  Jude computes the structural similarity~\cite{Wang2004ImageQA} for different prefix sizes over a sample of images, and uses this to derive a new utility function (e.g., \Cref{f:utility}). 

  \stitle{Improve the Predictor:} Jude now uses her application expertise to incrementally improve the predictor.  One direction is to weigh the point distribution with a prior based on historical image access frequency.  Alternatively, she could directly estimate the user's mouse position using a variety of existing approaches~\cite{pasqual2014mouse,wobbrock2009angle,wobbrock2007gestures,guo2008exploring,aydemir2013user}. She can assess the benefits of any modifications to the predictor based on its empirical accuracy over live or historical user traces, or higher-level metrics such as cache hit rates and number of blocks available for each request---\sys reports both.

\smallskip\noindent
In \myref{ss:exp_falcon}, we describe how we adapted the state-of-the-art Falcon DVE application to \sys with  $<100$
LOC.

\section{Predictor Manager}\label{s:predictor}

The application-provided prediction model $P^t(q | \Delta, e_t)$ uses
interaction events and/or requests $e_t$ up until the current time $t$ in order
to estimate the probability of request $q$ at $\Delta$ time steps in the
future.  Of course, there exist a wide range of prediction models that satisfy
this definition, with the appropriate one varying on a per-application basis.
For example, button and click-based interfaces benefit from Markov-based
models~\cite{he2009web,begleiter2004prediction,debrabant2015seer}, whereas
continuous interactions such as mouse- or hover-based applications benefit from
continuous estimation
models~\cite{aydemir2013user,welch1995introduction,pasqual2014mouse,wobbrock2009angle}.
Regardless of the prediction model used, a commonality with respect to \sys is
that the events $e_t$ (e.g., mouse movements, list of previous user actions)
are generated on the client, whereas the predictions are used by the
server-side scheduler.

Given these properties, \sys provides a generic API for applications to
register their desired predictors; \sys is agnostic to the specific prediction
model being suggested. The API (described below) decomposes a predictor into client-side and
server-side components, and \sys's Predictor Manager handles the frequency of
communication between the two components.  The main requirement is that the
predictor is usable at any time to estimate a probability distribution over
possible requests at arbitrary time steps. We note that \sys does not mandate a
specific prediction accuracy. However, \sys can report prediction accuracies,
as well as application-level performance metrics resulting from those
accuracies, based on live and historical user traces; developers can then use
this feedback to continually improve their predictors.

\stitle{Predictor decomposition.}  Applications specify the predictor $P^t$ 
as \red{server} and \blue{client}  components (correspondingly colored):
$$P^t(q|\Delta,e_t) = \red{P^t_s(q|\Delta,s_t)}\blue{P^t_c(s_t |\Delta,e_t)}$$
The client component $\blue{P^t_c}$ collects user interaction events and
requests $e_t$ and translates this information into a byte array that represents the predictor state $s_t$. 
$s_t$ may be the most recent request(s), model parameters, the most recent user
events, or simply the predicted probabilities themselves.  The server uses
$s_t$ as input to $\red{P^t_s}$ in order to return future request probabilities
for the \sys scheduler's joint optimization between prefetching and response
tuning.

Importantly, this decomposition is highly flexible and can support a variety of
different configurations for predictor components. For example, a pre-trained
Markov model~\cite{he2009web,begleiter2004prediction,debrabant2015seer} may be
instantiated on the server as $\red{P^t_s}$, and the client may simply send
each event to the server ($s_t = e_t$).  Alternatively, the Markov model could
be placed on the client as $\blue{P^t_c}$, with the state sent being a list of
the top $k$ most likely requests, and the server component assuming that all
non-top $k$ requests have probability of $\approx 0$\%.

\stitle{Devising A Custom Predictor.}
We now walk through the design of a custom predictor for interfaces with static
layouts, i.e., the two example DVE applications in \Cref{f:vis}.
These are the predictors
that we use in our experiments in \myref{s:experiments}. We note
that the purpose here is to elucidate the operation of an anytime predictor and
the process that a developer may follow in designing a suitable (custom) one
for their application. We do not claim that the following predictor is the best
possible one for a given application.

The two DVE applications in \Cref{f:vis} both use a fixed set of static
layouts: one uses a grid of thumbnails, while the other uses a set of
fixed-size charts.  Since requests are only generated when a user's mouse is
positioned atop a widget, the mouse position ($x,y$) is a rich signal for
predicting future requests. More specifically, the bounding boxes of all
widgets in the current layout $l$, denoted $\red{P_l}$, can directly translate
a distribution of mouse locations $\red{P^t_s(x,y|\Delta,s_t)}$ into a
distribution over requests:
$$P^t(q|\Delta,e_t) =
\red{P_l(q|\Delta,x,y,l)P^t_s(x,y|\Delta,s_t)}\blue{P^t_c(s_t,l|\Delta,e_t)}$$
  We model $\red{P^t_s(x,y|\Delta,s_t)}$ as a gaussian distribution represented by the centroid 
  and a $2\times2$ covariance matrix---this state is sent to the server.  
  We choose a fixed set of $\Delta$ values ($50, 150, 250, 500ms$ in our experiments) to predict over, and linearly interpolate between these times.
  Thus, the state $s_t$ only consists of $6$ floating point values for each $\Delta$, 
  which we estimate using a naive Kalman Filter~\cite{welch1995introduction} on the client, 
  and decode into a request distribution on the server.

Although it may appear challenging to devise a custom predictor, we note that
{\it any} cloud application that wishes to use prefetching will need to develop
or adapt a predictor. Further, our results in \myref{s:exp_khameleon} show that
the predictor need not be perfect, as \sys is effective {\it in spite} of the
generic Kalman Filter described above. Indeed, the fundamental challenge that \sys
solves is in determining how to explicitly and robustly account for the
predictions in its joint scheduling problem.

\section{Scheduler}\label{s:scheduler}

\sys's server-side scheduler takes as input a utility function $U$ and a
probability distribution over future requests, and
allocates finite network bandwidth and client cache resources across
progressively encoded data blocks to maximize the expected user utility.
Ultimately, it balances competing objectives: ensuring high utility for high
probability requests and hedging for lower probability requests (i.e., sending
some blocks for a low-quality response).

Developing this scheduler is challenging for several reasons. First, the
scheduler must keep track of previously sent blocks and ensure that they are
not evicted from the client's circular buffer cache by the time they are
needed.  Second, the scheduler needs to make decisions in real-time in order to
not block data transmission, but still must adjust its scheduling decisions
quickly when new predictions arrive from the client.  This section presents the
formal scheduling problem description, an ILP-based solution, and a greedy
approximation.

\begin{figure}[t]
    \centering
    \includegraphics[width=\columnwidth]{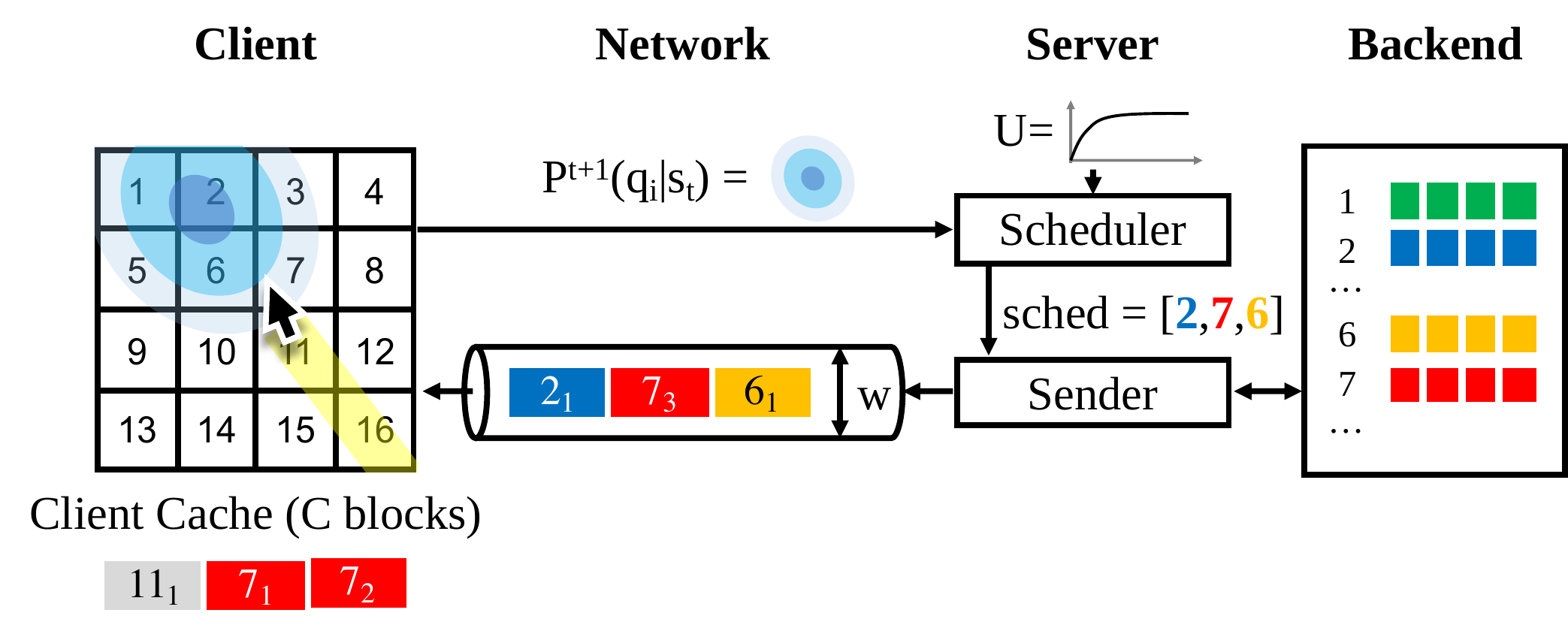}
    \caption{Setting for \sys's scheduling problem.}
    \label{fig:scheduling}
\end{figure}

\subsection{Problem Definition}

Let time be discretized such that each time interval $[t, t+1)$ is the time
that it takes for the server to add one block of a response onto the network.
In this problem definition, we assume that each response is progressively
encoded into $N_b$ equal-sized blocks.  

Let $Q = {q_1, \dots, q_n}$ be the set of all possible requests.
In \Cref{fig:scheduling}, there are $n=16$ possible image requests with ids $1$ to $16$.
The gaussian parameters estimated at time $t$ are the state $s_t$.
The scheduler has received predictor state $s_t$, which lets it estimate the  
probability $P(q_i | \Delta, s_t)$   of $q_i$ being issued at $\Delta$ time steps in the future.  
Let us assume that at the start of scheduling, $t=0$.

The client cache can hold $C$ blocks, and the network bandwidth is $w$ blocks
per time interval.  The cache at time $t$ contains $B_i^t$ blocks for $q_i$.
In the example, $2_1$ refers to the first block in image $2$.
The cache holds the first block of image $11$ ($B_{11}^t=1$), and the first two blocks of $7$ ($B_7^t=2$).
Thus, $B^{t+1} = \{B_1^{t+1}, \dots, B_n^{t+1}\}$ is the allocation at the end of the interval $[t, t+1]$.

\begin{problem}[Server-side Scheduling]
  Find the best next allocation $B^{t+1}$ that maximizes $V(B^t)$,
  given the cache $B^t$ and predictor state $s_t$: 
\begin{equation}
  \small
\label{eqn:schedule}
\begin{aligned}
  &V(B^{t+\Delta}) =  \\
  &\max_{B^{t+\Delta+1}}\left\{ \blue{\sum_i U(B_i^{t+\Delta+1})P(q_i|\Delta+1,s_t)}  + 
  \red{\gamma V(B^{t+\Delta+1})} \right\}
\end{aligned}
\end{equation}
\end{problem}

\noindent Our objective function $V$ includes two terms (colored in formula and text).   \blue{The first term is the expected user utility at the next timestep $t+\Delta+1$.  It weighs the utility of $B^{t+\delta+1}_i$ blocks (using the utility function) for request $i$ by its probability.}  \red{The second term recursively computes the future benefits.    This enforces the dependency between time intervals---it accounts for the long term and steers the scheduler towards a global optimum.}  
\red{$\gamma\in[0,1]$ is a discount on the future.  $\gamma=0$ means we only care about the next timestep, and $\gamma=1$ means we care about all timesteps equally.    } 

In \Cref{fig:scheduling}, the scheduler computes the best allocation for the next three time steps as the requests $2, 7, 6$.  The client cache's deterministic replacement policy lets the sender push the appropriate block sequence $2_1$, $7_3$, then $6_1$.

\subsection{Approximate ILP Scheduler}
Equation~\ref{eqn:schedule} is intractable because it contains an infinite recursion, as well as terms that are unknown at prediction time $t$.  However, due to the design of the client cache as a circular buffer, the cache will overwrite itself after every $C$ blocks.  Thus, we approximate the solution by optimizing over a finite horizon of $C$ blocks:
{\small\begin{multline}
\label{eqn:approx}
V(s_t, B^t) =
\underset{B^{t+1}, \dots, B^{t+C}}{max}
\sum_{k=1}^{C} \left(\gamma^{k-1}\sum_{i=1}^n U(B_i^{t+k})P(q_i,k)\right)
\end{multline}}
This formulation is a Markov Decision Process~\cite{Puterman1994MarkovDP}, where actions (chosen block) in each state (contents of the cache) receive a reward (user utility).  We now describe an ILP-based scheduler, followed by a fast real-time approximation.  In \myref{s:conclusion}, we discuss the relationship with reinforcement learning and future extensions.

\stitle{Objective Function.}
Equation~\ref{eqn:approx} can be expressed as an integer linear programming (ILP) problem.
ILP problems require a linear objective, but the utility function $U$ could be arbitrarily concave. We linearize it by approximating $U$ with a step function $\tilde{U}$, defined such that $\tilde{U}(0) = 0$ and $\tilde{U}(b) = \sum_{i=1}^{b} g(i)$ where:
{\small\[
    g(i) = U\left(\frac{i}{N_b}\right) - U\left(\frac{i-1}{N_b}\right)\ \mid\ i\in [1, N_b]
\]}
This approximation has no impact on the final result because  $U$ is already
discrete due to discrete block sizes.

Let $U_{i,j}^{t}$ denote the expected utility gain of the $j$-th block for
$q_i$ sent during time interval $[t-1, t]$, where $t\in[1,C]$. Because this
block is guaranteed to stay in the client cache until timestep $C$, it will
provide a constant amount of utility gain through time interval $[t, C]$.
Note that we dropped $s_t$ from $P$ since, from the perspective of the scheduler, it is a fixed prediction.  
{\small\[
U_{i,j}^{t} = \sum_{k=t}^{C} \gamma^{k-1} P(q_i|k) g(j)
\]}
 
We denote by $f_{i, j}^t$ a binary variable that indicates if the $j$-th block of $q_i$ is sent at time interval $[t-1, t]$. With this notation, we can transform the objective into a linear function:
{\small\begin{align}
\sum_{k=1}^{C} \left(\gamma^{k-1} \sum_{i=1}^n U(B_i^{k})P(q_i|k)\right)
= \sum_{i=1}^n \sum_{j=1}^{N_b} \sum_{k=1}^{C} f_{i,j}^k U_{i,j}^k
\end{align}}

\stitle{Constraints.} Our ILP program must account for three constraints. 
The ring buffer's limited capacity is implicitly encoded in
our choice of maximum time horizon $C$ in the objective.  
The ILP hard constraints ensure that (1) the network bandwidth is not exceeded, 
and that (2) each block is only sent once:
{\begin{align*}
    \forall k \sum_{i,j} f_{i,j}^k &\leq l &      %
    \forall i,j \sum_k f_{i,j}^k &\leq 1 %
\end{align*}}

\stitle{Limitations.} 
The LP scheduler is very slow because the LP problem size, 
as well as the cost to compute the utility gain matrix $U^t_{i,j}$, 
increases with the time horizon (cache size), the interaction
space (number of possible requests), and the granularity of the progressive
encoding (number of blocks).  
For instance, if the image application (10k possible requests) has a cache size of 5k blocks, and 10 blocks per request, 
the LP will contain $0.5$ billion variables.
Simply constructing the problem in the appropriate format for a solver is too expensive for a real-time system, and further,
this cost is incurred for every $C$ blocks to be sent.
\myref{a:sched-lp} reports our  micro-experiments, including comparisons with the fast, greedy scheduler described next.

\subsection{Greedy Scheduler}\label{ss:greedy}

This subsection describes \sys's fast greedy scheduler (\Cref{l:greedy}).  
The main design consideration is that it can rapidly make scheduling decisions 
as the client sends distributions at unpredictable rates, 
and without blocking (or being blocked by) the sender.
We first describe the scheduler design, and then discuss the interaction between the scheduler and the sender.  
\myref{a:scheduler-formal} describes the formal semantics of a correct schedule, given a sequence of distributions sent from the client.

\subsubsection{Greedy Scheduler Design}

Our greedy scheduler uses a single-step horizon (\blue{first term in \Cref{eqn:schedule}}). 
It computes the expected utility gain for giving one
block to each request (accounting for the number of blocks that have already
been scheduled), and samples a request $q_i$ proportional to its utility
gain.  The next block is allocated to $q_i$.  It schedules  batches of
$C$ blocks to fully fill the client cache, then it resets its state and repeats.

\stitle{State.}
The algorithm keeps three primary pieces of state that vary over time.  
The first is the number of blocks assigned to each request $B = [b_1,\ldots,b_n]$.  
This is used to estimate the utility gain for the next scheduled block, and is reset after a full schedule ($C$ blocks) have been scheduled.
The second state materializes $g()$ as an array.
The third state precomputes the matrix
$\mathbb{P}_{i,t} = \int_{k=t}^{C-1} P(q_i|k)$ that represents the 
probability that the user will request $q_i$ over the rest of the batch.  
This is estimated as a Reimann sum via the Trapezoidal Rule (lines 8-11).

Scheduling is now a small number of vectorized operations. 
The expected utility gain at timestep $t$ is the
dot product $\mathbb{P}_{t}\bullet g[B]$, where $\mathbb{P}_{t} =
[\mathbb{P}_{1,t},\ldots,\mathbb{P}_{n,t}]$ and $g[B] =
[g(b_1),\ldots,g(b_n)]$ are vectorized lookups (line 16).

\stitle{Scheduler Algorithm.}
The client is allowed to send new probability distributions at arbitrary times.
If a new distribution arrives, we wish to use its more accurate estimates, but also do not
wish to waste the resources used for prior scheduling work.
Further, the scheduler should progress irrespective of the rate at which the client sends distributions.

To make progress, each iteration schedules up to a batch of \texttt{bs} blocks at a time (default of 100).
After each batch, it checks whether a new distribution has arrived,
and if so, recomputes the $\mathbb{P}_{i,t}$ matrix (lines 6-11).  
Since \texttt{t} blocks may already have been chosen for the current schedule, 
we only need to materialize the time slots for the rest of the schedule ($\mathbb{P}_{i,t'}$ where $t'\in[t+1, C-1]$).
After sending the scheduled blocks to the sender, it resets \texttt{t} and \texttt{B} if a full schedule
has been allocated (lines 21-23).

\begin{figure}
\begin{lstlisting}[
  caption={Pseudocode of the greedy scheduler algorithm.},
  captionpos=b,
  label={l:greedy},
  basicstyle=\small\ttfamily,
  frame=,
  mathescape=true,
  numbers=left,
  escapeinside={@}{!}
]
C, g, n   @\grey{// cache size, utility array, \# requests}!
bs = min(C,bs)  @\grey{// blocks to schedule per iter}!
B = [0,..,0]    @\grey{// \# blocks for each req in cache}!
t = 0           @\grey{// \# blocks scheduled}!
while True:
  if received_new_distribution()
    dist = get_new_distribution()
    for i$\in$[1, n]
      $\mathbb{P}_{i,C}$ = dist(i, C)
      for t'$\in$[C-1, t]
        $\mathbb{P}_{i,t'}$ = $\frac{1}{t'+1}\mathbb{P}_{i,t'+1}$ + $\frac{t}{t'+1}$dist(i,t')

  S = [ ]       @\grey{// generated batch of blocks}!
  while t < C-1 and |S| < bs
    t += 1
    u = $\mathbb{P}_t \bullet g[B]$
    q = sample requests proportional to u
    S.append(q)
    B[q] += 1

  send S to sender
  if t == C     @\grey{// reset after a full schedule}!
    t,B = 0, [0,..,0]
\end{lstlisting}
\end{figure}

\stitle{Optimizations.}  
We employ several straightforward optimizations beyond the pseudocode in \Cref{l:greedy}.
The main one avoids materializing the full $\mathbb{P}_{i,t}$ matrix when the number of requests is very high.
Most requests will have the same probability of $\approx 0$
(images 4, 8, 9, 12, 13-16 in \Cref{fig:scheduling}), and correspondingly similar utility gains.
Thus, we group these requests into a single ``meta-request'' whose probability is the sum of the individual requests.
If the scheduler samples this meta-request in line 17, then it uniformly chooses one of the individual requests.
On a benchmark with 10K requests, 5K blocks in the cache, and 50 blocks per request, this optimization reduces the cost of generating one schedule from $1.9s$ to $150ms$ ($13\times$ reduction).  
Using this concept to further stratify the probability distribution may further reduce runtime, but we find that this optimization is sufficient in our experiments.

\subsubsection{Sender Coordination}
Our current prototype assumes that the client and server clocks are
synchronized to ensure that servers can ensure sufficient confidence in
predictions, and that the Sender thread can be preempted.  When a new
prediction arrives at the scheduler, it identifies the position $i$ of the
sending thread in the current batch, and reruns the scheduler from $i$ to $C$.
The blocks for $0$ to $i$ do not change since they have already been sent.  
This is analogous to setting $t=i$.

The scheduler sends this partial schedule to the sending thread, which in the
meantime, may have sent an additional $h$  blocks.  Thus, it simply starts
sending using the partial schedule at $i+h$.  Concurrently, the scheduler
begins scheduling the next batch using the updated predictions.  Note that the
scheduler may be modified to match a different client-cache replacement
strategy; we leave this to future work.

\subsection{Implementation Details}
\label{ss:impl}
\stitle{Bandwidth Estimation.} The sender thread and scheduler require
knowledge of the available network bandwidth, and aim to run at a rate that
will not cause congestion on the network. \sys is agnostic to the specific
end-host bandwidth estimation (and prediction) technique that is used to
compute this information~\cite{Yin2015ACA,proteus,pytheas}. Further, note that
\sys can alternatively be configured to use a user-specified bandwidth cap
(e.g., to comply with limited data plans). In our implementation,
the \sys client library periodically sends its data receive rate to the server;
the server uses the harmonic mean of the past 5 rates as its bandwidth estimate
for the upcoming timestep, and aims to saturate the link. This approach
capitalizes on recent observations that bandwidth can be accurately estimated
over short time scales, particularly in backlogged settings that avoid
transport layer inefficiencies (e.g., TCP slow-start-restart)~\cite{pensieve}
that mask the true available bandwidth at the application
layer~\cite{Yin2015ACA}.

\stitle{Backend Scalability.}
This work assumes that backend query execution is scalable, i.e.,
data stores can execute many concurrent speculative requests without performance
degradation.  This is often true for key-value-oriented backends~\cite{wu2019anna} or cloud
services, but may not hold for other data stores. For instance, databases such
as PostgreSQL have a concurrency limit, after which per-query performance
suffers.  Thus, it is crucial for the scheduler to avoid issuing too many
speculative requests such that the backend becomes a bottleneck in response
latency.

Although formally addressing this interaction challenge between \sys and data
processing backends is beyond the scope of this paper, we use a simple heuristic to
avoid triggering congestion in the backend. We benchmark the backend offline to
measure the number of concurrent requests $C$ that it can process scalably. 
Let $n$ be the number of requests the backend is currently processing; we
post-process schedules to ensure that they do not refer to
blocks from more than $C-n$ distinct requests.  In essence, we treat backend
request limits in the same way as network constraints.

\section{Experiments} \label{s:experiments}

We evaluate \sys on the DVE applications described
in \myref{s:usecases}.
Our experiments use real user interaction traces and
a wide range of network and client cache conditions. We
compare against idealized classic prefetching and response tuning
approaches, highlight the benefits provided by each
of \sys's features, and investigate \sys's sensitivity to 
environmental and system parameters. 
  The results comparing \sys with the baselines are consistent across the applications.
Thus, we primarily report results for the image application, and use Falcon to illustrate how \sys goes well beyond the state-of-the-art hand-optimized implementation (\myref{ss:exp_falcon}).

\begin{figure}[t]
  \centering
  \includegraphics[width = .9\columnwidth]{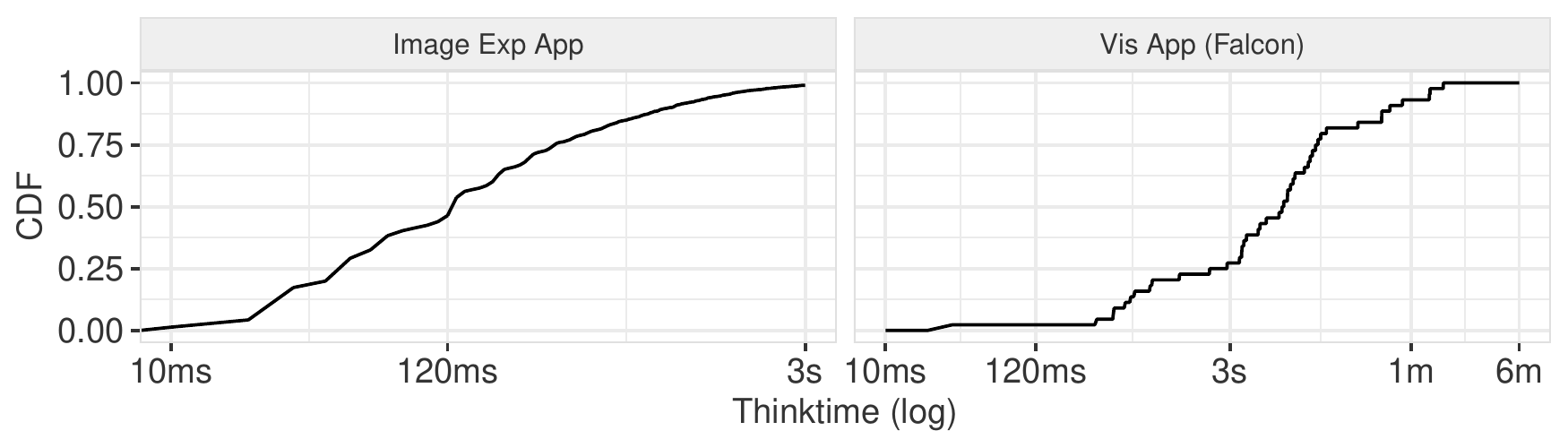}
  \caption{CDF of think times (time between consecutive requests) over interaction traces for the image and vis applications (\myref{s:usecases}).}
  \label{f:traces_mosaic}
\end{figure}

\subsection{Experimental Setup}
\label{s:exp-setup}

Our prototype uses a Typescript client library and a Rust server library.  
The client periodically sends predictions to the server every $150ms$.
Each prediction consists of a distribution over the possible requests at
timesteps $\Delta=\{ 50, 150, 250, 500ms\}$ from the time that the prediction is
made; the $500ms$ values follow a uniform distribution. The application backends
precompute and progressively encode results for all possible requests.

We use the image exploration and visualization applications described in
\myref{s:usecases}.
For the image application, we collected mouse-level interaction traces from 14 graduate students that
freely used a version of the application that was configured with no response latency. Each trace is 3 minutes long, with $20ms$ average think time.  
For Falcon, we used the 70 traces from~\cite{battle2020database}. The interface used to collect these traces differs from the interface in the Falcon paper~\cite{moritz2019falcon} by one chart (a bar chart instead of the heat map in~\cite{moritz2019falcon}).  Thus we translated the interactions over the bar chart to generate semantically equivalent requests consistent with~\cite{moritz2019falcon}.  In this way, the performance numbers are comparable with~\cite{moritz2019falcon}.
We find that increasing the number and length of traces doesn't affect our results; \Cref{f:traces_mosaic} reports the think-time distributions.

\stitle{Performance metrics:}
\sys balances response latency and quality for preemptive interactions.
However, due to the bursty nature of interactions, some requests (and their response data) may be preempted when later requests receive responses sooner.  Thus, we report the percentage of preempted requests.  For the non-preempted requests, we measure the cache hit rate as the requests that have blocks in the cache at the time of the request,  the response latency as the time from when a request is
registered with the cache to the upcall when one or more blocks are cached, and the response utility at that time.  We will also evaluate how quickly the utility is expected to converge to 1 (all blocks).

\iffalse
The above quality and latency tradeoffs are nuanced and multi-faceted.  To simplify comparisons between settings, we propose a new metric called {\it responsiveness} $r_\lambda = \lambda u + (1-\lambda)\hat{l}$ that is a combination of the response utility $u$ and normalized latency $\hat{l}$. $\lambda\in[0,1]$ balanaces the weight between the two terms, and the normalized latency is $1$ when the latency $l$ is below 100ms, and decays to $0$ as latency increases (\Cref{f:normlat}):  
$\hat{l}= \begin{cases}
  1 & if\ l\le100\\
  e^{-\frac{x-100}{500}} & else
\end{cases}$ 
\fi

We use the utility curves in Figure~\ref{f:utility}.
The image application's utility function is described in \myref{ss:archcomponents}.   
Falcon implements progressive encoding by sampling rows of the response in a round-robin fashion. For instance, for a 1D CDF, we sample values along the x-axis. We conservatively use the default linear utility function.

\stitle{Environment parameters:} 
Our experiments consider a wide
range of network and client-side resource scenarios. 
We first use netem~\cite{netem} to consider fixed bandwidth values between $1.5$--$15MB/s$ \footnote{We report the bandwidth as \texttt{MB/s} instead of \texttt{Mb/s} to use the same units as block sizes.}
 (default $5.625MB/s$) and request latencies between $20$--$400ms$ (default
$100ms$); note that because we precompute all responses, request latency is
meant to include both network latency (between $5$--$100ms$) and simulated backend
processing costs ($15$--$300ms$). We vary the client's cache size between $10$--$100MB$ (default $50MB$).  We also use the Mahimahi network emulator~\cite{mahimahi} to emulate real Verizon and AT\&T LTE cellular links; in these experiments, the minimum network round trip time was set to $100ms$~\cite{watchtower}.   We simulate varying think time between requests from $10$--$200ms$, which is favorable to the baseline approaches described below.   \Cref{f:traces_mosaic} shows CDFs of think times in our user traces.

\stitle{Performance baselines:}
 \baseline is a standard request-response application with no prefetching.  \prog mimics \baseline, but only retrieves the first block of any response---this is intended to reduce network congestion but does not use prefetching to mask request latency.

Prefetching techniques primarily focus on prediction accuracy and the number of parallel requests to make.  Modern predictors exhibit $\le70\%$ accuracy~\cite{battle2016dynamic}.  To create strong baselines (\texttt{ACC-<acc>-<hor>}), we use a perfect predictor that knows the next \texttt{hor} requests with \texttt{acc} accuracy per request. After each user-initiated request, the prefetcher issues up to \texttt{hor} prefetching requests; to avoid triggering network congestion, it does not prefetch if the number of outstanding requests will exceed a bandwidth-determined threshold.  For example, after the $i^{th}$ user request, \texttt{ACC-.8-2} will predict the $i+1^{th}$ and $i+2^{th}$ requests, and each will have 80\% chance of being correct (i.e., matching the actual request in the trace).  We use \texttt{ACC-0.8-1}, \texttt{ACC-1-1}, and \texttt{ACC-1-5} (following~\cite{battle2016dynamic}).  All baselines use an LRU cache.

We also evaluate an \oracle version of \sys where the predictor knows the exact position of the mouse after $\Delta$ milliseconds (by examining the trace).

\subsection{Comparison with Baselines}
\label{s:exp_baseline}

We first compare \sys with the aforementioned baselines including no prefetching, and \texttt{ACC-0.8-1}, \texttt{ACC-1-1}, and \texttt{ACC-1-5}. Recall that these are upper bounds for existing prefetching approaches---typical predictors have accuracies of $<70\%$.

\begin{figure}[t]
  \centering
  \includegraphics[width=\columnwidth]{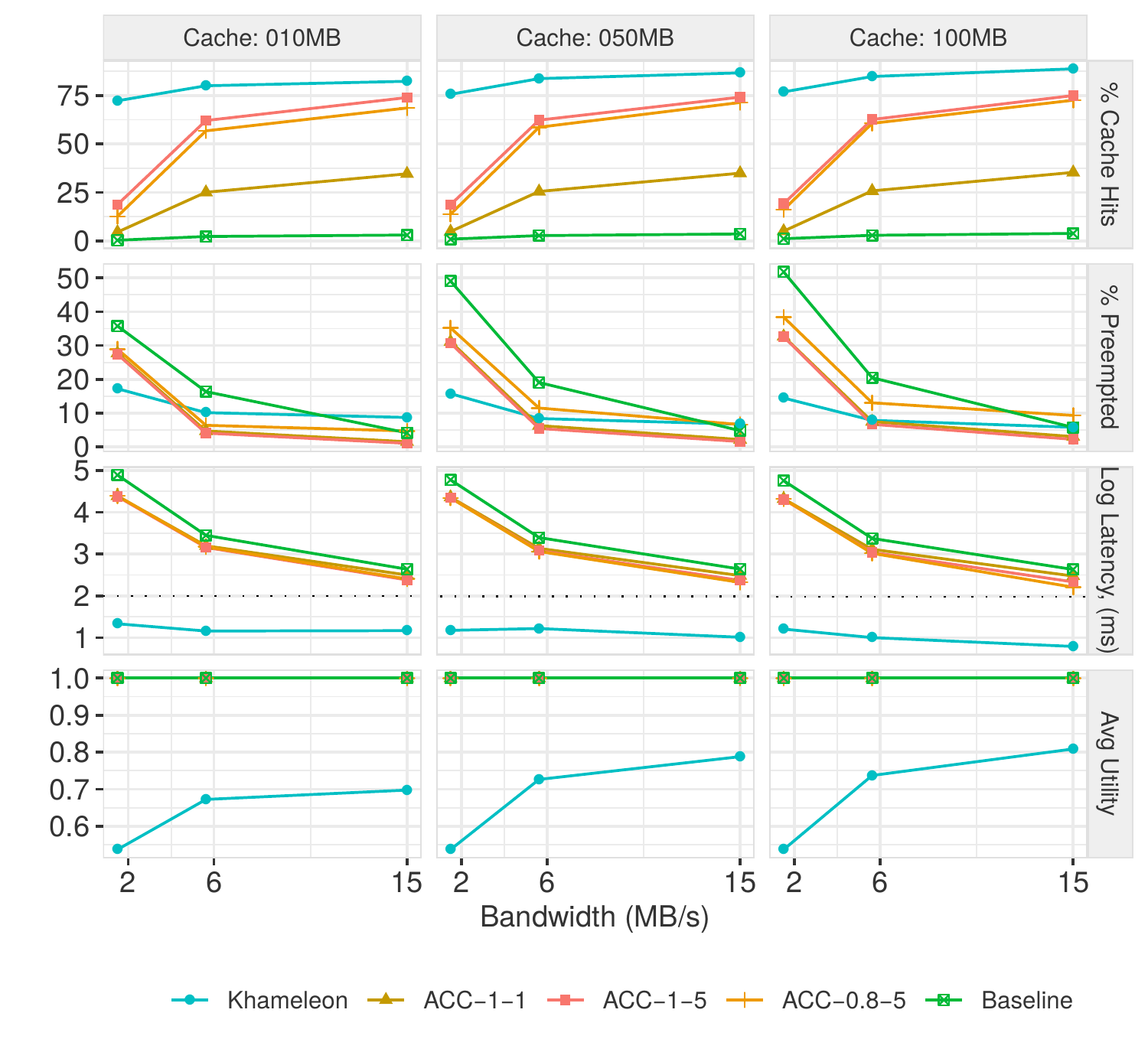}
  \caption{Idealized prefetching baselines and \sys across varying cache and network bandwidths (x-axis). Pane headers list the metric for each y axis.  Latency charts in all figures render a black dashed line at $100ms$.}
  \label{f:exp-resource-vis}
\end{figure}

\stitle{Varying Bandwidth and Cache Resources: }
We faithfully replayed the user traces, and varied the cache size ($10$--$100MB$)
and bandwidth resources ($1.5$--$15MB/s$), while keeping request latency fixed to 100 ms.  
The top two rows in Figure~\ref{f:exp-resource-vis}
report the percentage of requests for which one or more blocks are present in the
cache at the time of request (i.e., \% Cache Hits), and the percentage of preempted requests. 
\sys increases the cache hit rate by $23.38-256.73\times$ above \baseline, and by $1.11-16.12\times$
above the idealized prefetching baselines (\texttt{ACC-*-*}). 
\sys reduces the number of preempted requests by $3\times$ in low bandwidth settings, 
and has slightly higher preemption rate than  \texttt{ACC-*-*} at higher bandwidths because its high cache hit rate causes more out-of-order responses.
The \texttt{ACC-*-*} baselines have lower cache hits because think times are lower than request latency, thus the user has moved on by the time the prefetched data arrives.

The bottom two rows plot the utility and user-perceived response latency
for requests that are {\it not preempted}.  We see that the
baselines consistently prioritize full responses---their utilities are
always $1$ at the expense of very long response latencies (note latencies are log scale).  In contrast, \sys
gracefully tunes the utility based on resource availability---all the while 
maintaining consistently low average response
latencies that do not exceed 
$14ms$ across the bandwidth and cache size configurations.
\textbf{On average, across different cache resources and bandwidth limits, \sys has up to $16\times$ better cache hit rates than \texttt{ACC-*-*}, resulting in $16.35$--$1525.23\times$ lower response times.}

\begin{figure}
  \centering
  \includegraphics[width=\columnwidth]{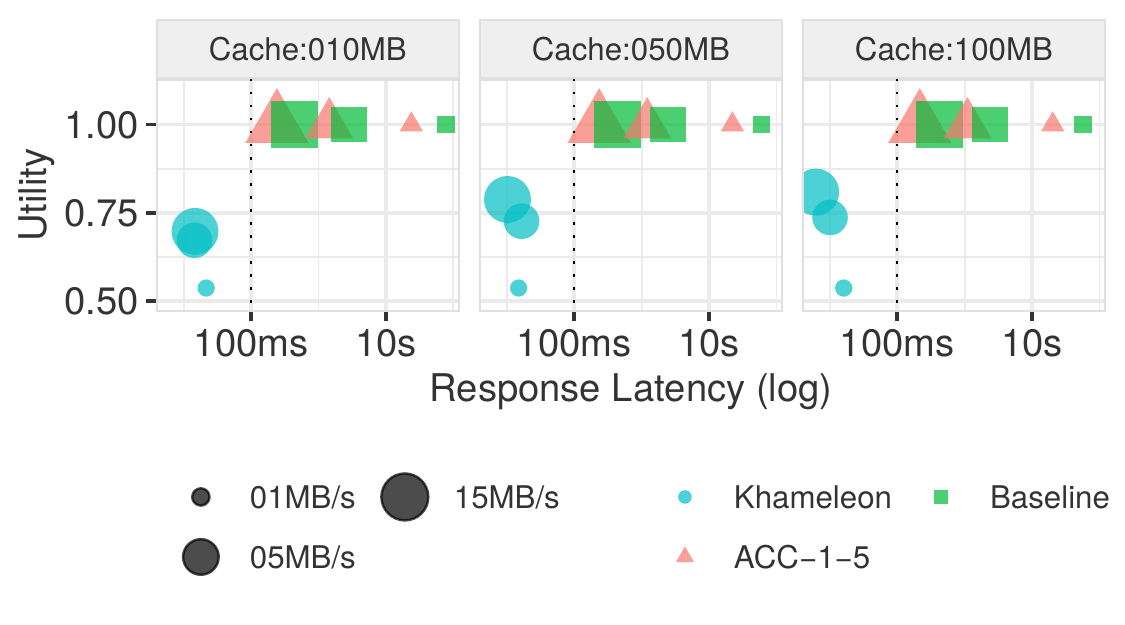}
  \caption{Response time vs utility for the prefetching baselines and \sys. Size denotes bandwidth; upper left is better.  Black dotted line shows $100ms$ latency threshold.}
  \label{f:exp-resource-lat-v-util-vis}
\end{figure}

To better illustrate the tradeoff between resources, responsiveness, and utility, \Cref{f:exp-resource-lat-v-util-vis} compares average
response latency (across all requests) and the response utility, for every
condition (shape, color), bandwidth (size), and cache size; upper left means faster response times and higher utility. Across all conditions, increasing the bandwidth improves the response times. However, the baselines remain at perfect utility and have high latencies. In contrast, \sys always has $<100ms$ latency and judiciously uses resources to improve utility.

\begin{figure}
  \centering
  \includegraphics[width = \columnwidth]{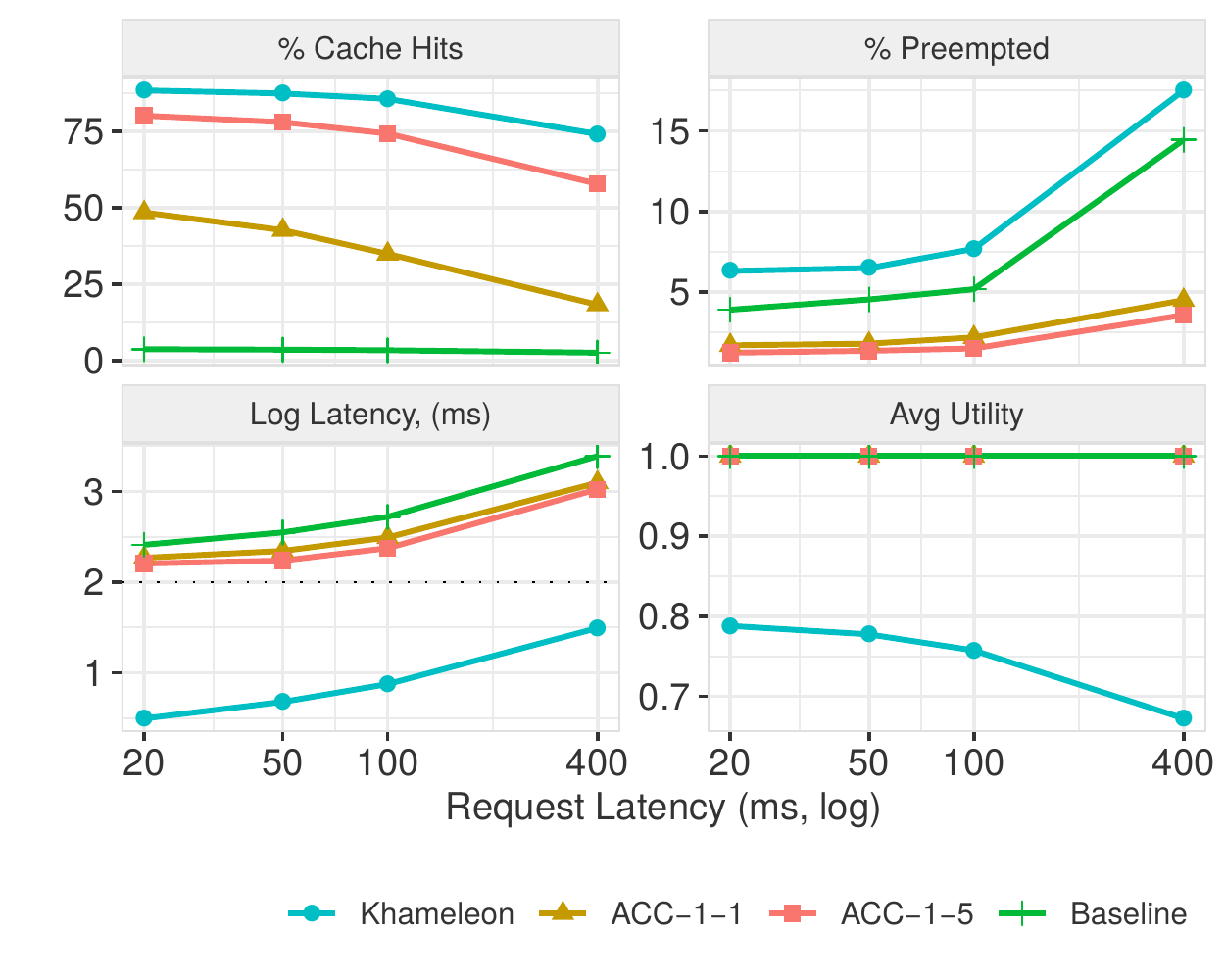}
	\caption{\sys vs prefetch baselines across varying request latencies; request latency includes both network and server
processing delays.}
        \label{f:latency_sweep}
\end{figure}

\stitle{Request latency:} We now fix network bandwidth ($15MB/s$) and cache size ($50MB$),
and vary request latency ($20$--$400ms$). Recall that request latency includes both
network and server processing delays. 
Figure~\ref{f:latency_sweep} shows that \sys consistently achieves higher cache hit rates
than the prefetching baselines.
As request latencies grow, \sys degrades response
utility to ensure that response latencies remain low (on average $11ms$)
and have on average $3\times$ higher preempted requests than baselines because of the out-of-order responses that results from higher cache hit rate.
In contrast, the alternatives pursue perfect utilities at the detriment of responsiveness.
When the request latency is $400ms$, Khameleon performs $79\times$ faster than Baseline,
and $37\times$ than \texttt{ACC-*-*}.
The baselines become highly congested as the request latency increases.

\begin{figure}[t]
  \centering
  \includegraphics[width=\columnwidth]{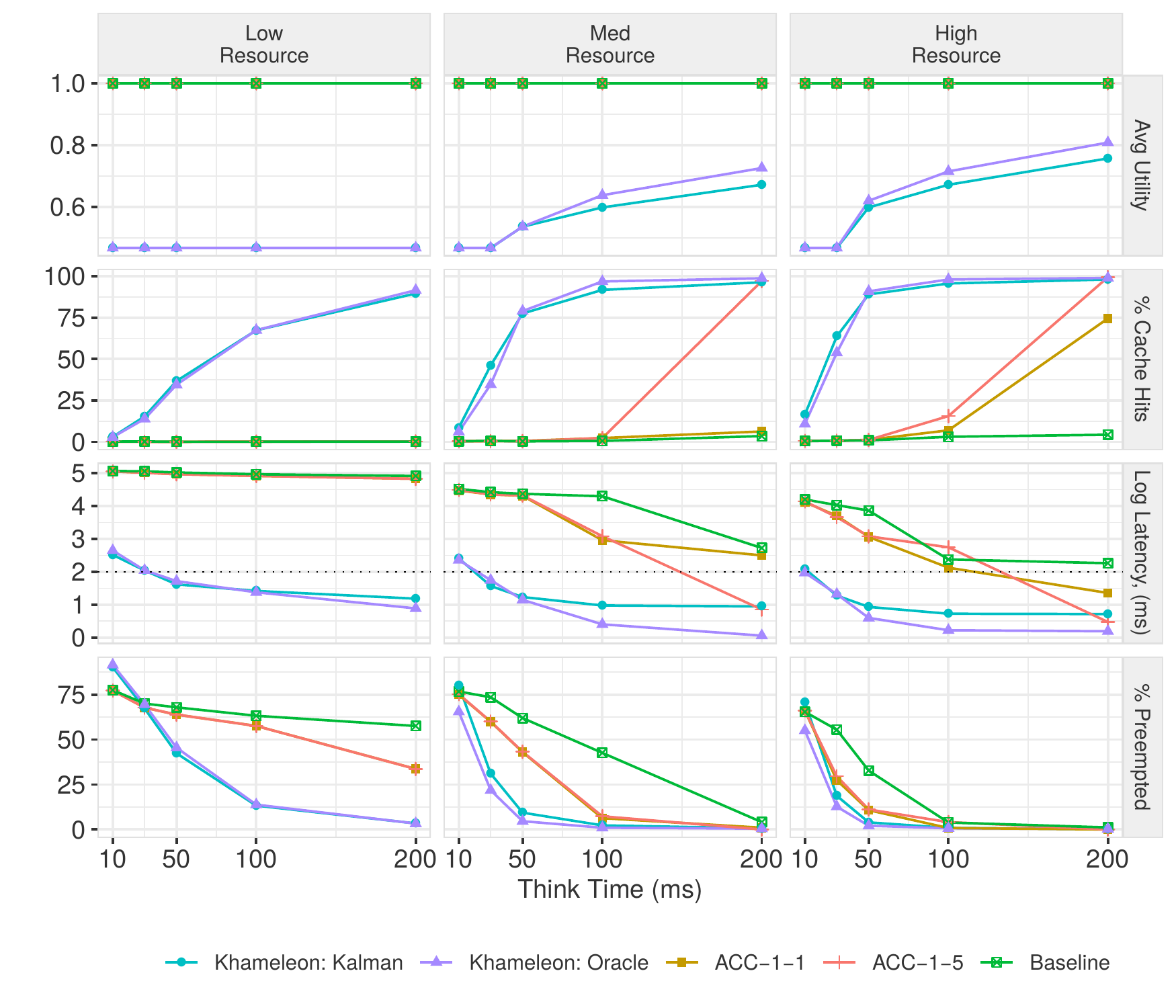}
  \caption{Varying think time between consecutive requests.  Comparing \sys vs \texttt{ACC} using prefect and kalman filter predictors for 1 and 5 request horizons. }
  \label{f:exp-thinktime-vis}
\end{figure}

\stitle{Think time:} So far, we have faithfully replayed the user traces.
Now we synthetically vary the think times in the traces between $10$--$200ms$ to assess its effect.  
We fix request latency to $100ms$, and use three resource settings:
low (bandwidth=$1.5MB/s$, cache=$10MB$), medium ($5.625MB/s$, $50MB$), 
and high ($15MB/s$, $100MB$).  

\Cref{f:exp-thinktime-vis} indeed shows that high think times improve all prefetching methods by reducing congestion and giving more time to prefetch.  
This is most notable in the high resource setting, where the \texttt{Baseline} response latency
converges to the cost of the network latency plus the network transmission time.
\texttt{ACC-1-*} has high response latency when the think time is short due to congestion,
but the cache rate increases to $75-100\%$ with high think time and high resources.  
With low resources and low think times, \sys achieves low latency by hedging, as shown by the low utility values.
Despite this, the next experiment shows that \sys converges to full utility faster than the baselines.
With more resources, \sys shifts to prioritize and improve utility.
We find that \sys is close to \oracle, except in high resource settings, where perfect prediction can better use the extra resources and further reduces latency by $2\times$.  

\textbf{\sys maintains near-instant response latencies, and uses the additional think time to increase the response utility.  }
This highlights \sys's efficacy for DVE applications with low think
times relative to request latency, i.e., where there is not enough time to
prefetch between requests, even with perfect prediction.

\begin{figure}
  \centering
  \includegraphics[width=\columnwidth]{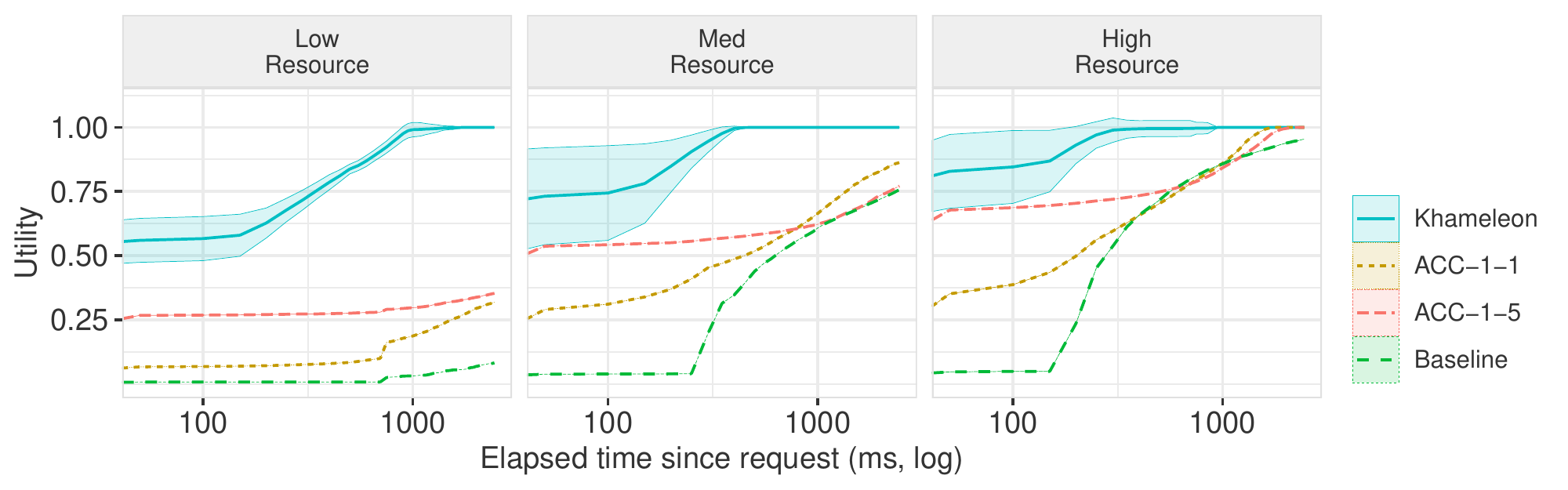}
  \caption{\small Convergence rate of utility:  average utility (y-axis) over time after the user stops on a request.   }
  \label{f:exp-converge}
\end{figure}

\stitle{Convergence:} 
Although trading utility for responsiveness is important, the
response should quickly converge to the full utility when the user pauses on a request. 
We now pause a user trace at a random time, and track the upcalls until the utility reaches 1.
We use the high, medium, and low resource settings described above.  
Figure~\ref{f:exp-converge} reports the average and standard deviation of the utility after waiting an elapsed time after pause.\footnote{\footnotesize The baselines always have utility of 0 or 1, so we only report the average.}
\sys consistently converges to a utility of 1 faster than all of the baselines, in expectation.
This is explained by the additional congestion incurred due to the high rate of requests issued by the two baselines.
We expect that better application-specific predictors~\cite{battle2016dynamic} can greatly improve convergence for \sys.

\subsection{Understanding Khameleon}
\label{s:exp_khameleon}

We now perform an ablation study, and vary system configurations to better understand \sys.

\stitle{Ablation Study. }
To show the importance of jointly optimizing resource-aware prefetching and progressive
response encoding, we perform an ablation study. Starting with a non-prefetching \baseline, 
we add the kalman filter and joint scheduler but without progressive encoding (\pred), 
and we add progressive encoding but without prefetching to show the benefits 
of cache amplification (\prog). For reference, we compare with \texttt{ACC-1-5}.
We use a bandwidth of 15MB/s, cache size of 50MB, and vary request latencies.  

\begin{figure}[t]
  \centering
  \includegraphics[width=\columnwidth]{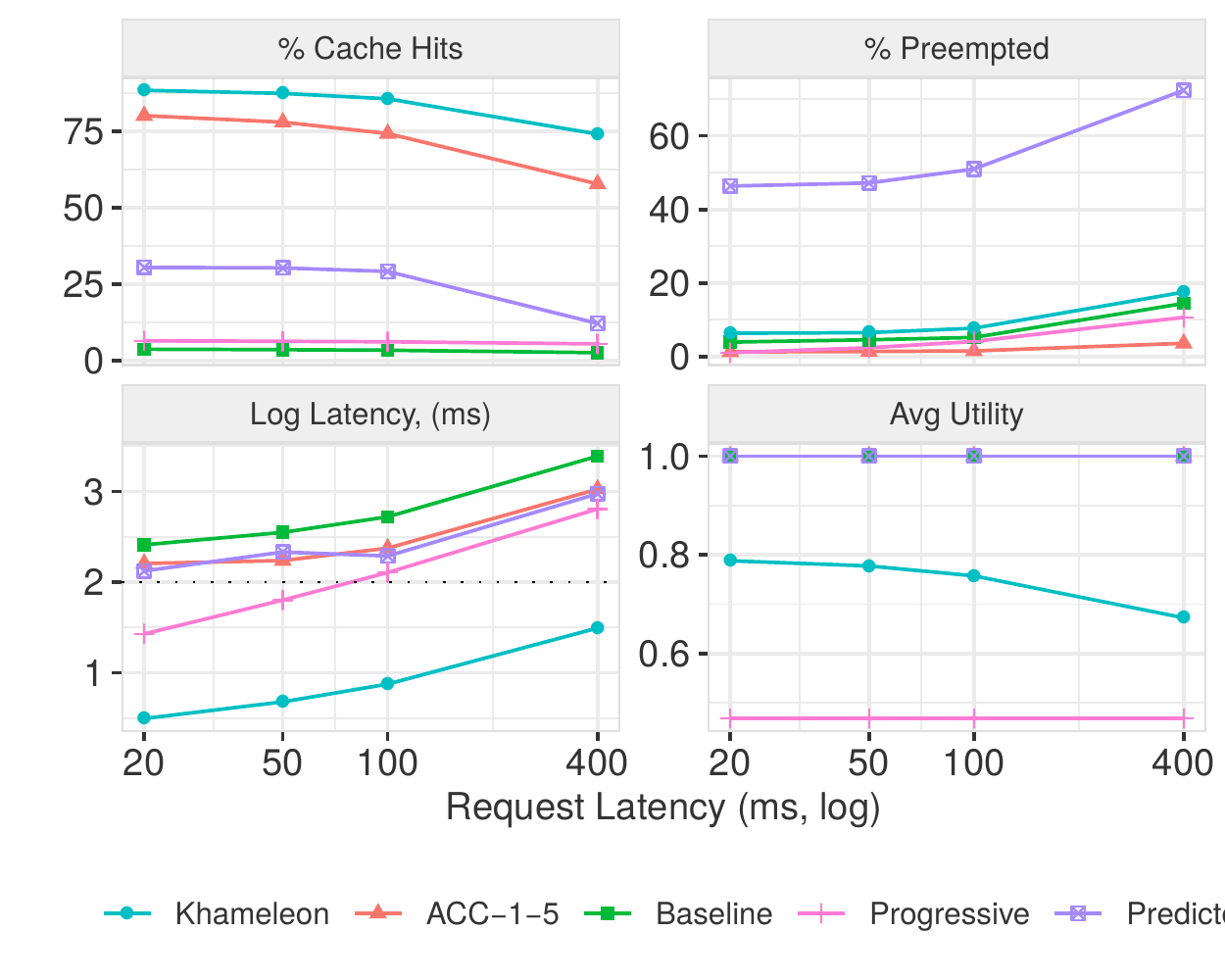}
  \caption{Results of ablation study.}
  \label{f:exp-breakdown-latency-vis}
\end{figure}

Figure~\ref{f:exp-breakdown-latency-vis} shows that as the request latency increases,
the cache hit rate for all approaches decreases (a negligible decrease for \sys). 
\pred improves over \baseline because the joint scheduler pushes predicted requests proactively (thus increasing the cache hit rate) without increasing network congestion.
\prog improves over \baseline by reducing the network transmission time and alleviating congestion,
yet its utility is also the lowest. 
The combination of the two optimizations are needed in \sys for higher utility, consistently $<31ms$ response, and ($>74\%$) cache hit rate.

\begin{figure}
  \centering
  \includegraphics[width=\columnwidth]{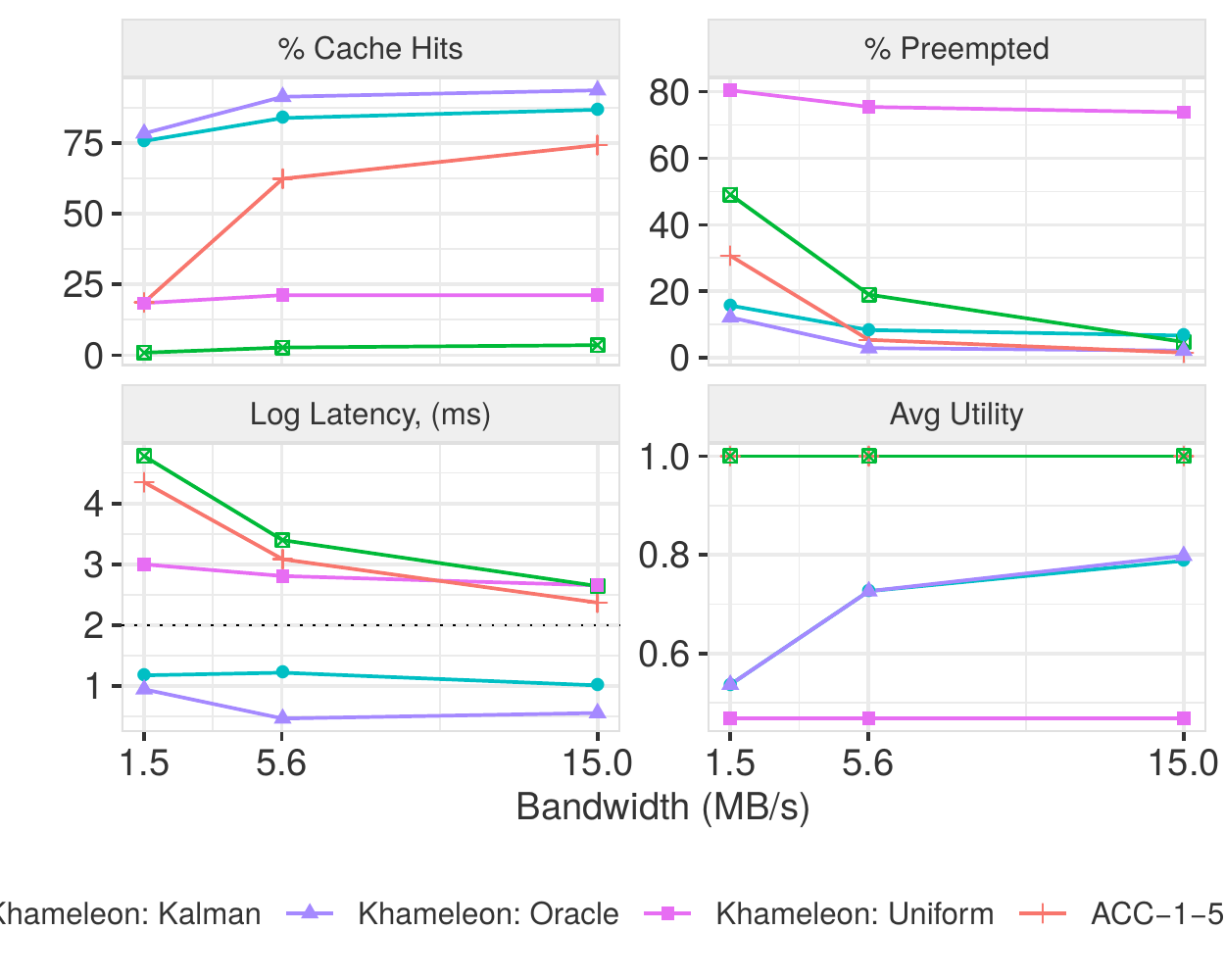}
  \caption{Varying \sys{} predictors.}
  \label{f:exp-predictor}
\end{figure}

\stitle{Sensitivity to Predictors. }
\Cref{f:exp-predictor} assesses the impact of the predictor by comparing the \uniform predictor, \kalman, and the \oracle predictor as the upper bound.
We fix request latency to $100ms$, and include \texttt{ACC-1-5} and \baseline as references.
At low bandwidth, simply using the \sys framework already improves latency compared to \texttt{ACC-1-5}; and \kalman further improves on top of \uniform and is close to \oracle.  As bandwidth increases, a more accurate predictor better uses the resources to push more blocks for more likely requests.  Thus, \oracle further reduces response times by $1.7-5.7\times$ compared to \kalman.

\stitle{System Parameters and Bandwidth Overheads. }
We evaluated the frequency that the prediction distributions are sent to the server, and find that \sys is robust to frequencies between $50-350ms$, but deteriorates when frequencies are lower.   We also measured the percentage of blocks that were pushed to the client but unused by the application (overpushed blocks): we find that \sys overpushes $50-75\%$ of the blocks, as compared to $35-45\%$ for \texttt{ACC-1-5}.  Since the user can limit the amount of bandwidth allocated to prefetching, we believe these rates are acceptable given the orders of magnitude lower latency. More details are in \myref{a:overhead}

\begin{figure}[t]
  \centering
  \includegraphics[width=\columnwidth]{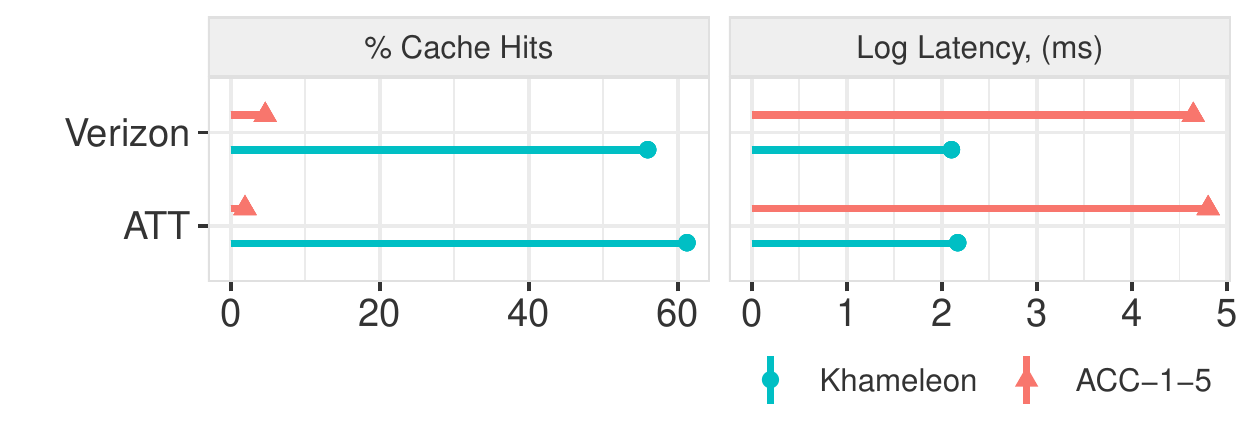}
  \caption{Comparing \sys with baselines on time-varying cellular networks.}
  \label{f:cell_traces}
\end{figure}

\stitle{Real Network Traces. }
On the real Verizon LTE, and AT\&T LTE cellular network traces, with a fixed $100ms$ request latency and $50M$ cache size, Figure~\ref{f:cell_traces} shows that \sys considerably outperforms \texttt{ACC-1-5}. 
The cache hit rate is over $10\times$ higher on AT\&T, and the latency is lower by $348.36-430.12\times$.

\subsection{Falcon Visualization Experiments}\label{ss:exp_falcon}

We now adopt Falcon~\cite{moritz2019falcon} to \sys, and show that the ability to easily change the predictor and introduce progressive encoding lets \sys further improve over the already--optimized Falcon.  

\stitle{Porting Falcon: }  We modified the Typescript client to register requests to the \sys client library.  Originally, when the user hovers over a chart, Falcon issues 5 separate SQL queries to the backend database to generate a data slice for each of the other five charts (we call this group of queries a single {\it request}).   We simulate this with a predictor that assigns a probability of 1 to the currently hovered upon view, and 0 to all others.   Similarly, when the scheduler allocates one block for a given request, the sender issues 5 queries to the query backend (PostgreSQL database), and progressively encodes the combined query results into blocks.   
In contrast to the image application, the backend only scales to 15 concurrent queries before query performance degrades considerably.  Thus, prefetching even 3 requests can issue enough queries to saturate the backend.

Adapting the client required $\approx50$ LOC---mostly decoding blocks into Falcon's expected format. The code to encode query results into blocks on the server required $\approx60$ LOC.

\stitle{Experiment Setup: }   We create two databases using subsets of the flights dataset from Falcon~\cite{moritz2019falcon}; \smalldb has 1M records with query latencies of $\approx 800ms$, and \bigdb has 7M records with latencies of $1.5$--$2.5s$.  We verified that the ported version performed the same as the original Falcon, so we report metrics based on varying the ported version.

\begin{figure}[h]
  \centering
  \includegraphics[width=\columnwidth]{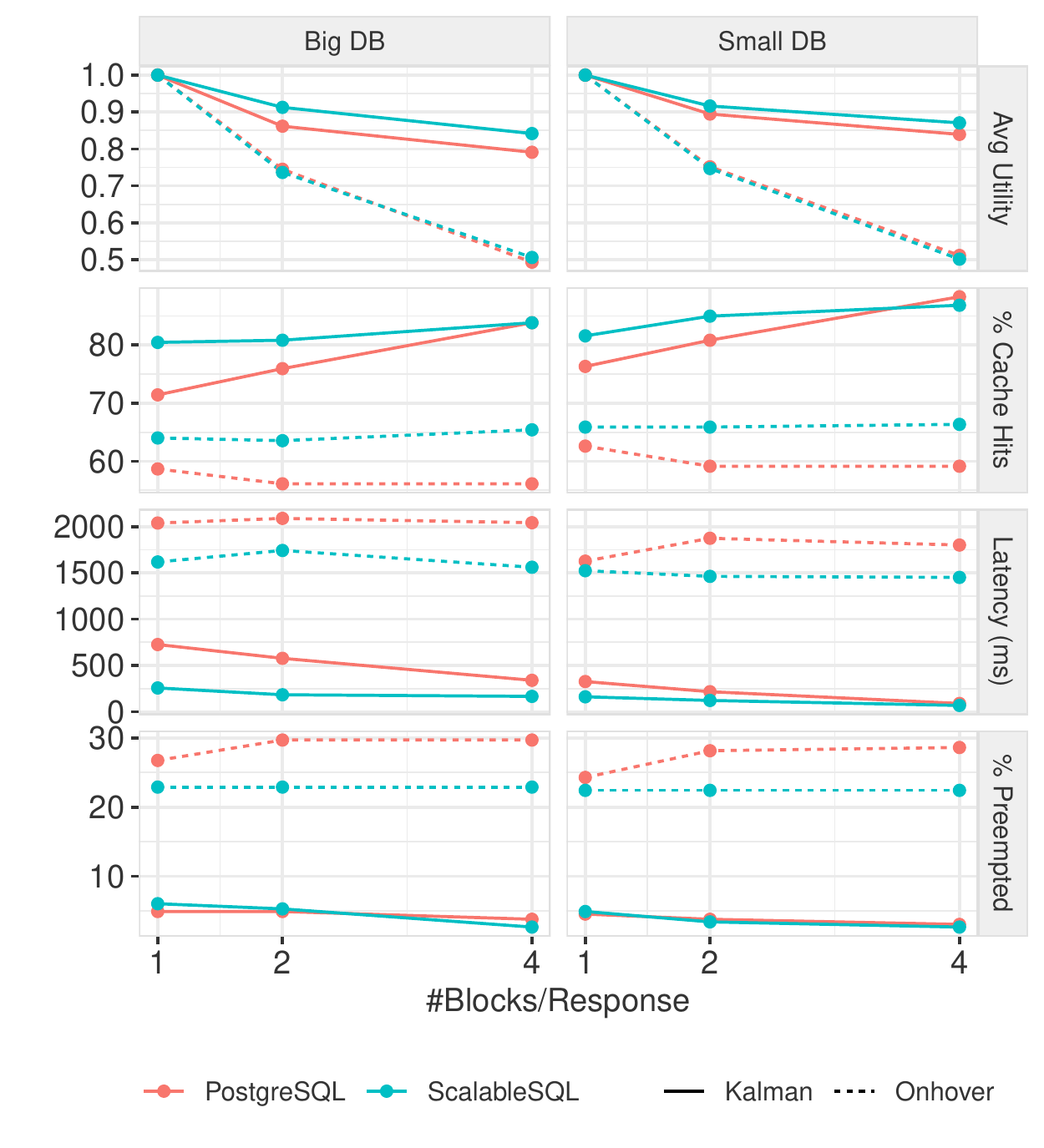}
  \caption{Ported Falcon system on \bigdb (7M) and \smalldb (1M) datasets with varying number of blocks/request (x-axis), and predictors (line) using Postgres (red line) and a simulated scalable database backend (blue line). }
  \label{f:falcon_exp}
\end{figure}

\stitle{Predictor and Progressive Encoding: }
We change the predictor from Falcon's ``on-hover'' (dashed lines) to the kalman filter (solid lines) used in earlier experiments. The x-axis varies the number of blocks that each request is encoded into (each block has fewer result records).
The red lines in \Cref{f:falcon_exp} (PostgresSQL) show that \kalman improves over \onhover, delivering $1.4\times$ more cache hits, $5\times$ lower latency on average, lower preemption rate, and higher utility across the two datasets, particularly as the number of blocks increases.

\stitle{Scalable Backend: }
We now simulate a scalable database (blue lines).  We first precompute and log each query's execution times when running in isolation.   The backend answers queries from a cache and simulates the latency.  Compared to the PostgreSQL backend,
\kalman response latencies improves on average by $2\times$, and \onhover by $1.2\times$. \kalman still outperforms \onhover with higher utility because it hedges more aggressively without congesting the backend.

\section{Related Work}\label{s:related}

\stitle{DVE application optimizations.}
Many existing approaches reduce DVE
application response latencies by addressing server-side data processing and
client-side rendering costs.  Query processing in backend datastores can be
completed in tens of milliseconds using a combination of precomputation (of
datacubes~\cite{Gray1995DataCA}, samples~\cite{ding2016sample},
indexes~\cite{el2016vistrees}), hardware~\cite{mapdimmerse,graphistry} or
vectorized~\cite{Zukowski2012VectorwiseBC} acceleration, and
parallelization~\cite{wu2019anna}.  Client-side computation and rendering
delays can be similarly reduced using techniques such as
datacubes~\cite{lins2013nanocubes} and GPU rendering and processing
acceleration~\cite{graphistry,liu2013immens,meyerovich2013superconductor}.

\sys is complementary to the above approaches and focuses on reducing the
network bottlenecks (not client- or server-side computation delays) in DVE
applications.  Indeed, \sys could be used as the communication layer to
progressively encode and push optimized data structures and query results based
on anticipated user interactions. Our current implementation makes the
simplifying assumption that data processing and progressive encoding incur a
fixed cost; leveraging the above query processing and response encoding
optimizations will require incorporating data processing costs into the
scheduler, a promising future direction.

\stitle{Caching and Prefetching. } \label{s:prefetch}
Interactive data exploration has studied various caching architectures~\cite{Sisneros2007AMC}, as well as prefetching techniques to pre-populate the caches~\cite{chan2008maintaining,doshi2003prefetching,ahmed2012adaptive,battle2016dynamic}.
ATLAS~\cite{chan2008maintaining} extrapolates scrollbar movements to prefetch
subsequent requests; ForeCache~\cite{battle2016dynamic} extends this to
prefetching tiles of array data by forecasting likely future queries based on
past user actions and statistical features of the underlying data;
These approaches are crafted for their specific visualization interface, and leverage restrictions of the user's allowed interactions (often to linear slider actions or panning to nearby tiles) that help improve the prediction accuracy. 
Falcon~\cite{moritz2019falcon} prefetches when a user hovers over a chart, and uses the time between hovering and interacting with the chart to prefetch datacube structures.  
Database query prefetching typically relies on Markov models that update when a new query is issued~\cite{ramachandran2005dynamic,sapia2000promise,dice2014,cetintemel2013query}, which assumes longer think times between queries, while web page requests~\cite{domenech2006web,nanopoulos2003data,white2017search} use mouse predictors similar to the Kalman Filter used in the experiments.

Though these techniques are able to perform backend query computation early,
they do not incorporate server push mechanisms or progressive response
encoding, limiting their impact on alleviating network bottlenecks. \sys
borrows similar prediction information, but replaces explicit requests and
responses with probability distributions and a fine-grained scheduler for
push-based streaming that accounts for request probabilities \textit{and}
response quality.

\stitle{Progressive Encoding.} 
Progressive encoding ensures that a small prefix
of the overall response data is sufficient to render an approximate result,
while scanning additional blocks improves result quality (ultimately converging
to the fully accurate version)~\cite{hoppe1996progressive}.  This has been
applied to a wide range of data, including
images~\cite{salomon2010handbook,shapiro1993embedded}, layered
encodings~\cite{Jakubczak2011ACD,Segall2007SpatialSW,Ghandi2006H264LC,Dardari2004LayeredVT},
visualization data~\cite{battle2016dynamic}, and even web page
resources~\cite{netravali2018vesper,netravali2016polaris,ruamviboonsuk2017vroom}.
\sys lets applications provide progressively encoded responses~\cite{salomon2010handbook,shapiro1993embedded,battle2016dynamic}, which enables the scheduler's joint optimization to dynamically trade off response quality for low latency in DVE applications. 

\stitle{Progressive Computation.}
Online aggregation~\cite{hellerstein1997online,agarwal2012blink,li2016wander,alabipfunk,rahman2017ve} and progressive visualization~\cite{fekete2015progressivis,moritz2017trust,fisher2012trust,stolper2014progressive,loadngo} seek to quickly return approximate results whose estimates improve over time, and could be used as backends in \sys.
DICE~\cite{Kamat2014DistributedAI} also uses speculation to accelerate visualization exploration.  DICE bounds the query space to faceted data cube navigation,  speculatively executes approximate queries for neighboring facets across a sharded database, and allocates sampling rates to the queries based on the expected accuracy gains.

These progressive computation techniques are fundamentally different than the \emph{progressive encoding} techniques considered by \sys. In progressive computation, each improvement is {\it a separate result set} that requires sending more data and using more network resources.  The progressive encoding with \sys could be used to encode a result set.  Thus, although both techniques achieve a similar {\it effect} of progressively enhancing the rendered data, the mechanisms are different and complementary.

\pagebreak
\section{Discussion and Conclusion}
\label{s:conclusion}

\sys is a dynamic prefetching framework for data visualization and exploration
(DVE) applications that are approximation tolerant.  Rather than focusing
solely on predicting requests to prefetch or adapting response quality to
available resources, \sys uses a server-side scheduler to jointly optimize
across these techniques.  Responses are progressively encoded into blocks, and
proactively streamed to the client cache based on request likelihoods.  

\sys consistently achieves sub-30ms response times even when requests take
400ms, and out-performs existing prefetching techniques (often by OOM).  It
gracefully uses resources to improve quality.  To best leverage \sys,  each
component in the system (the backend scalability, network bandwidth, degree of
speculative prefetching) should be matched to produce and consume data at the
same rates.

  \stitle{Learning Improved Policies.}
  This work used unsophisticated prediction models and scheduling policies, rather than sophisticated models or policies, to argue the effectiveness of a continuous push framework.  As expected, we also found a considerable gap from an optimal predictor; we expect better scheduling policies as well.  One extension is to adapt a Reinforcement Learning framework to improve the scheduler's policy. For instance, we could log explicit reward signals in the client, and use Q-learning to better estimate future rewards.  We could also unroll beyond a single step, and use policy gradient methods~\cite{Sutton1999PolicyGM} to learn a higher quality policy function that may account for deployment idiosyncrasies.  The challenge is to balance more sophistication with the need to schedule the next block in real-time (microseconds).

\stitle{Acknowledgements:} Thanks to Dan Rubensein, Adam Elmachtoub for advice on the ILP formulation; Thibault Sellam, Mengyang Liu on early system versions; NSF IIS 1845638, 1564049, 1527765, and CNS-1943621.

{
\bibliographystyle{abbrv}
\bibliography{main}
\pagebreak
}

\begin{appendix}
\section{Scheduler Details}\label{a:scehduler}

\subsection{Microexperiments for LP and Greedy Schedulers}
\label{a:sched-lp}

We implemented the LP-based scheduler in Rust
using Gurobi, a state-of-the-art linear program (LP) solver~\cite{jablonsky2015benchmarks}.

\begin{figure}[h]
  \centering
  \includegraphics[width=\columnwidth]{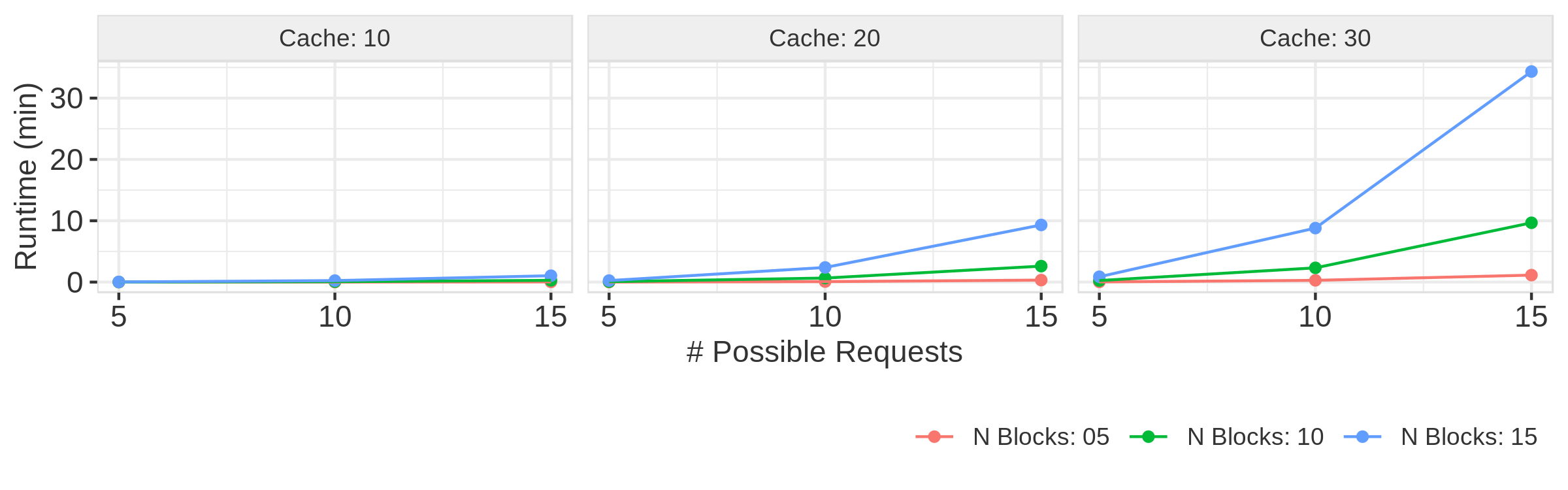}
  \caption{LP scheduler runtime}
  \label{fig:exp_sched_runtime_ilp}
\end{figure}

Figure~\ref{fig:exp_sched_runtime_ilp} reports the runtime when varying the number of possible
requests between 5-15, the cache size from 10-30 blocks, and the
number of blocks per request from 5-15.  
The LP scheduler is very expensive, as the size of the LP increases proportionally with the
number of possible requests, the cache size, and the number of blocks.
Even for such trivial scenarios, generating a schedule can take up to 30 minutes.

\begin{figure}[h]
  \centering
  \includegraphics[width=\columnwidth]{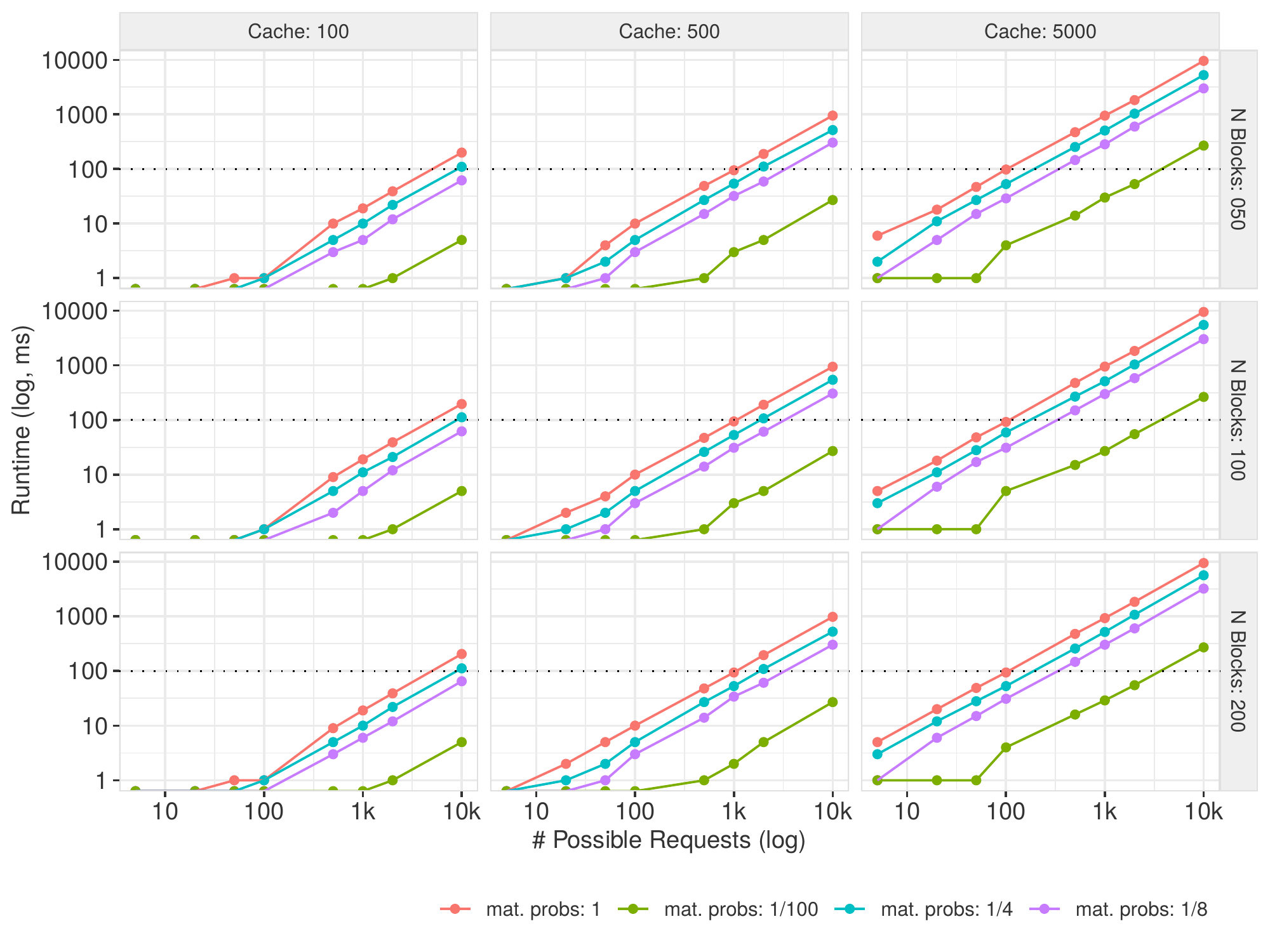}
  \caption{Runtime of the greedy scheduler across varying cache size, number of requests, number of blocks per request (N Blocks), and the percentage of the requests that requires to be materialized (mat. probs) in the matrix $\mathbb{P}_{i,t}$}

  \label{fig:exp_sched_runtime_greedy}
\end{figure}

Figure~\ref{fig:exp_sched_runtime_greedy} shows the performance of the
(optimized) greedy scheduler as the cache size, number of blocks, and number of
possible requests vary; in particular, the figure lists the scheduler's runtime
and the fraction of requests that have non-uniform probabilities and thus must
be materialized. As shown, the running time is independent of the number of
blocks and it increases proportionally with the number of requests and cache
size.  The bulk of the cost is in precomputing the probability matrix
(\Cref{l:greedy}, lines 8-11).  Consequently, the running time of the scheduler
is heavily dependent on the fraction of requests with non-uniform
probabilities.
From our experience, this fraction is low for many DVE applications. For example, for the image gallery application with 10k requests, the fraction was less than $\frac{1}{100}$.
Moreover, the greedy algorithm offers flexibility to configure the batch size to produce a schedule in realtime ($<100ms$).

\begin{figure}[h]
  \centering
  \includegraphics[width=\columnwidth]{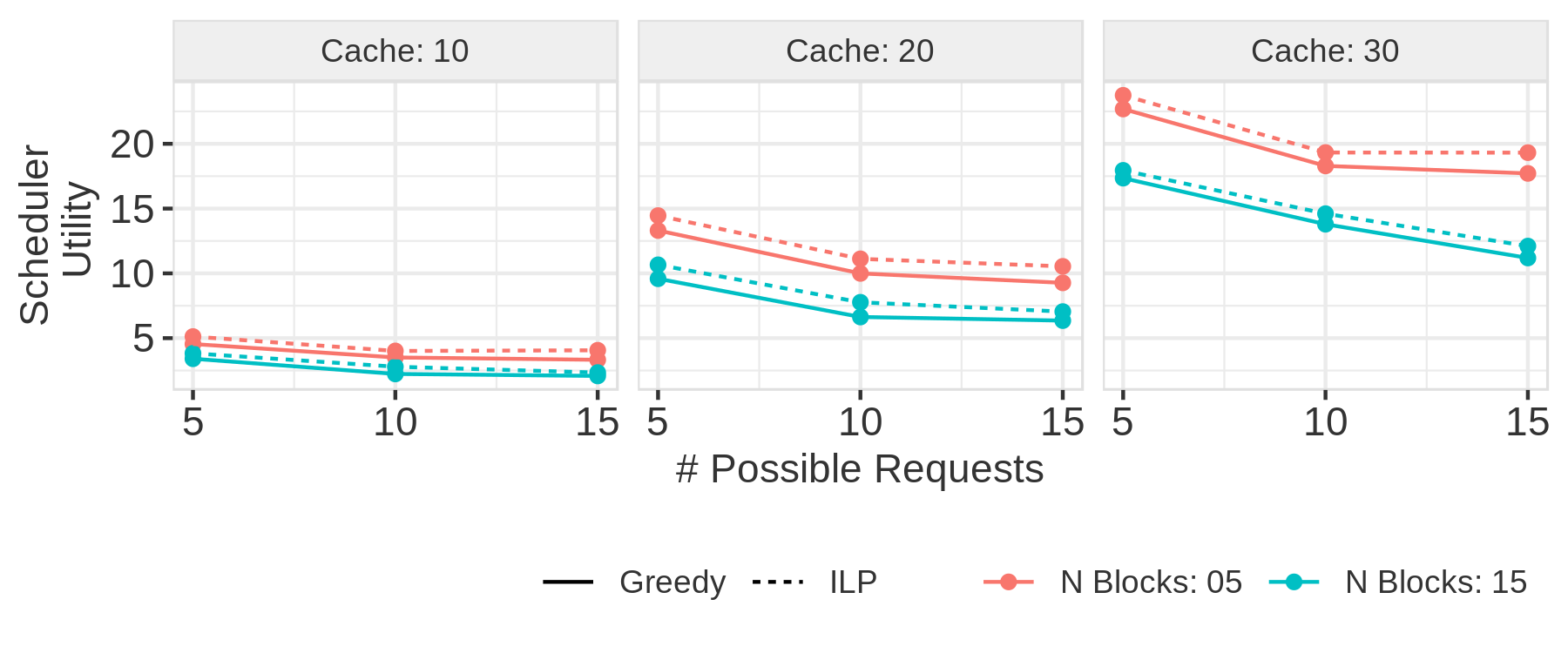}
  \caption{Greedy and LP schedule utilities.}
  \label{fig:exp_sched_utility_ilp_greedy}
\end{figure}

Further,
\Cref{fig:exp_sched_utility_ilp_greedy} shows that the schedules generated
using the greedy approach have competitive utility as compared to the LP
scheduler (on average $1.2\times$ less than LP scheduler), 
while benefiting from a $\ge3000\times$ reduction in runtime.

\subsection{Details of Multiple Prediction Distributions}
\label{a:scheduler-formal}

\begin{figure}[t]
  \centering
  \includegraphics[width = .8\columnwidth]{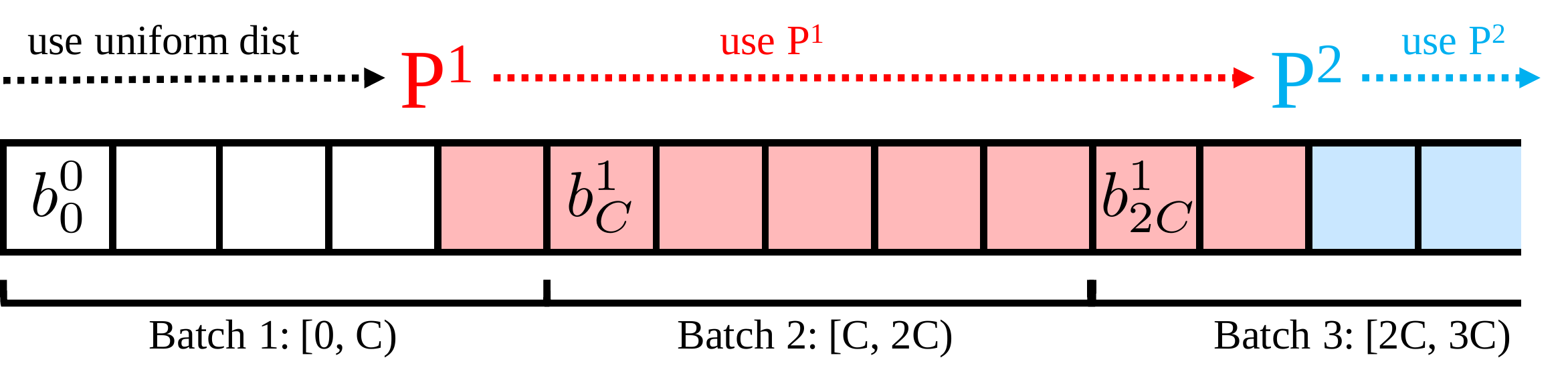}
  \caption{Semantics of idealized scheduler as new predictions $P^j$ arrive. The blocks have the same color hue as the prediction they rely upon. }
  \label{f:scheduler-example}
\end{figure}

This subsection describes the formal semantics of a schedule when the client sends
a sequence of prediction distributions to the scheduler.

Formally, consider an analysis session of $n$ timesteps, where each timestep
corresponds to the time to place one block on the network.  Suppose the
one-way network latency is fixed as $l$, and the client sends $k$ predictions
to the server, where prediction $P^j()$ is sent at time $t_j-l$ and arrives at
the server at $t_j$.  Let the global schedule be $s=[b_1,\dots,b_n]$ where the
server simply sends each block in sequence.  Let $b^j_i$ be the block that the
scheduler would pick for timestep $i$ if it used $P^j$ to schedule the block,
under the following conditions:
\begin{itemize}[leftmargin=*, topsep=0mm,itemsep=0mm]
\item if $i<t_1$, then assume a uniform probability distribution,
\item if $i<t_j$, then $b^j_i$ is null,
\item if $i\ge t_j$, then $b^j_i$
is computed as part of batch $m=\floor*{\frac{i}{C}}$ (e.g., timesteps $[mC,
(m+1)C]$), and is the $i-mC$ block in the batch.  \end{itemize}

\noindent Given this, $b_i = b^j_i$, where $j = max \{ t_j \le i | j\in [0,
k]\}$ is the most recent prediction to arrive prior to timestep $i$.  In
\Cref{f:scheduler-example}, the scheduler creates a schedule for the first
batch of $C=5$ blocks using a uniform distribution.  When $P^1$ arrives, it is
used to reschedule $b_4$ in the first batch, all blocks in batch 2, and blocks
in batch 3.   However, $P^2$ arrives before timestep $13$, so the rest of
batch 3 is rescheduled using $P^2$.  The superscript for each block denotes the
prediction that the scheduler used.  Naturally, this is an idealized setting
where there are no scheduler delays, network variance, and other timing
nondeterminism.

\section{Additional Experiments}\label{a:overhead}

\subsection{System Parameters}

We varied the {\it frequency} that the client library sends predictions to the server from
every $50$--$350ms$ across the low, medium, and high resource settings (recall that the default value used in our experiments thus far is 150ms).  Overall, varying the frequency has a minor effect on the reported metrics (cache hit rate, response latency, utility, and preemption rate) and negligible compared to the effects of other settings (e.g., network or cache resources).  The one exception is in the low cache size and low bandwidth setting, where sending predictions less frequently ($>300ms$) reduces the prediction accuracy and wastes precious resources on irrelevant data.

\subsection{Bandwidth Overheads} 
It is clear that prefetching makes a trade-off between bandwidth and responsiveness.  We measured the percentage of blocks sent by the server that were not involved in upcalls to answer application requests (called {\it overpushed} blocks).  We collected these statistics during the Think Time experiments.  %
 Overall, at most $75\%$ of the prefetched blocks by \sys are overpushed .  This is typically expected---particularly if the intention is to hedge across many possible future requests.  

\begin{figure}
  \centering
  \includegraphics[width=.7\columnwidth]{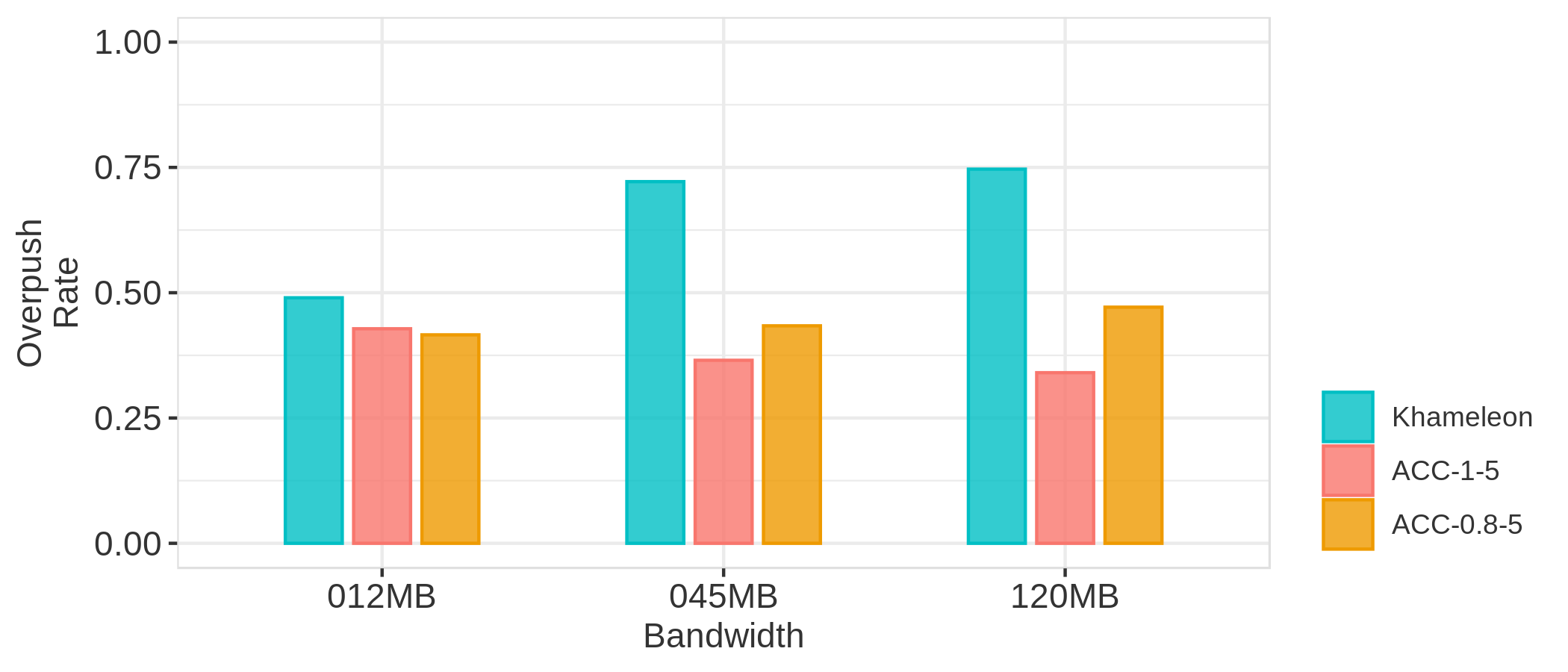}
  \caption{Overpush rate: percentage of data pushed to the client that were not used in an upcall to the application.}
  \label{f:exp-overpush}
\end{figure}

\textbf{A key benefit of \sys is that the user can decide how much bandwidth to allocate to prefetching, and the scheduler will scale back the amount of blocks scheduled and pushed accordingly.}    Additionally, the penalty of overpushing an incorrect block is lower because each block requires fewer network and cache resources than prefetching a full response.  

\end{appendix}

\end{document}